\documentclass[USenglish]{article}	
\usepackage[utf8]{inputenc}				%(only for the pdftex engine)
\usepackage[big,online]{dgruyter}	%values: small,big | online,print,work
\usepackage{lmodern} 
\usepackage{microtype}
\usepackage[numbers,square,sort&compress]{natbib}

\theoremstyle{dgthm}

\theoremstyle{dgdef}

\usepackage{amsmath}
\usepackage{graphicx}
\usepackage{hyperref}
\usepackage{physics}
\usepackage{todonotes}
\usepackage{booktabs}
\usepackage{url}
\usepackage[normalem]{ulem}
\usepackage{float}

\usepackage{amsmath}
\usepackage{bm}
\usepackage{multirow}
\usepackage[skip = 7.5pt, indent = 15 pt]{parskip}
\usepackage{subfig}
\usepackage{url}
\usepackage{appendix}

\usepackage{setspace}

\usepackage{csquotes}% Recommended

\definecolor{darkgreen}{rgb}{0.0, 0.5, 0.0}

\newcommand{\blind}{0}

\begin{document}

%%%--------------------------------------------%%%
\articletype{Research Article (\textit{Accepted for publication in Journal of Quantitative Analysis in Sports, 2026 soccer edition})}
	%\received{Month	DD, YYYY}
	%\revised{Month	DD, YYYY}
%  \accepted{Month	DD, YYYY}
  \journalname{De~Gruyter~Journal}
%  \journalyear{YYYY}
%  \journalvolume{XX}
%  \journalissue{X}
  \startpage{1}
 % \aop
 % \DOI{10.1515/sample-YYYY-XXXX}
%%%--------------------------------------------%%%

\title{Leveraging Minute-by-Minute Soccer Match Event Data to Adjust Team’s Offensive Production for Game Context}
\runningtitle{Leveraging Minute-by-Minute Soccer Data to Adjust Production for Game Context}
%\subtitle{Insert subtitle if needed}

\if0\blind{
\author*[1]{Andrey Skripnikov}
%\ use * to mark the author as the corresponding author
\author[2]{Ahmet Cemek, David Gillman}
%\author[2]{David Gillman} 
\runningauthor{Skripnikov et al.}
\affil[1]{\protect\raggedright 
Natural Sciences Division, New College of Florida, USA, e-mail: askripnikov@ncf.edu}
\affil[2]{\protect\raggedright 
Natural Sciences Division, New College of Florida, USA}
}
\fi

\if1\blind{
\author[1]{Blinded}
\affil[1]{\protect\raggedright 
Blinded}
}
\fi

%\communicated{...}
%\dedication{...}
	
\abstract{

In soccer, game context can result in skewing offensive statistics in ways that might misrepresent how well a team has played. For instance, in England’s 1-2 loss to France in the 2022 FIFA World Cup quarterfinal, England attempted considerably more shots (16 to France’s 8) and more corners (5 to 2), potentially suggesting they played better despite the loss. However, these statistics were largely accumulated when France was ahead and more willing to concede offensive initiative to England. To explore how game context influences offensive performance, we analyze minute-by-minute event-sequenced match data from 15 seasons across five major European leagues. Using count-response Generalized Additive Modeling, we consider features such as score and red card differential, home/away status, pre-match win probabilities, and game minute. Moreover, we leverage interaction terms to test several intuitive hypotheses about how these features might cooperate in explaining offensive production. The selected model is then applied to project offensive statistics onto a standardized "common denominator" scenario: a tied home game with even men on both sides. The adjusted numbers - in contrast to regular game totals that disregard game context - offer a more contextualized comparison, reducing the likelihood of misrepresenting the relative quality of play.
}


\maketitle

\section{Introduction}

In soccer, commonly referred to as football, statistics play an important role in analyzing various aspects of the game—from assessing individual player and team performances to providing valuable strategic insights. For instance, metrics such as shot attempts and corner kicks can indicate how offensively dominant a player or team was during a match. However, relying solely on aggregate statistics without considering the broader context of game events may lead to misleading conclusions.

Consider the 2022 FIFA World Cup quarterfinal between England and France (where "FIFA" stands for Fédération Internationale de Football Association). England recorded 16 shot attempts compared to France's 8, and earned 5 corner kicks to France's 2, yet still lost 1--2 \citep{espnFranceEngland}. Based on these numbers alone, one might conclude that England played a more dominant offensive match despite the loss. However, these raw totals ignore the context of the scoreline: during the 40 minutes when the match was tied, France led in the aforementioned categories—7 to 5 in shot attempts and 2 to 0 in corner kicks. In the remaining 66 minutes, with France in the lead, they purposefully ceded possession and initiative to England as a tactical decision, focusing their efforts on defense and protecting the lead.

The goal of this study is to develop a statistical adjustment that accounts for crucial contextual factors—such as score differential—when interpreting offensive metrics like shot attempts and corner kicks. In particular, we explore how various contextual variables affect offensive production, and subsequently project those effects onto a standardized "common denominator" scenario: a home game where the score is tied and both teams are at even strength throughout. By doing so, we aim for the statistical outputs to more accurately reflect teams' relative offensive performances, accounting for the likelihood of specific tactical adjustments under different game scenarios and reducing the risk of misleading conclusions. Additionally, leveraging minute-by-minute data allows us to test several hypotheses regarding how the impacts of various contextual factors may vary depending on the game minute—particularly when comparing early-game versus late-game situations.

In this section, we will first review existing metrics in soccer analytics, along with related research in other sports. We will then outline our research question and emphasize its contributions to the field.

    \subsection{Background and Literature Review}

The field of soccer analytics has been growing over the recent years, including the strong increase in scientific publications associated with it over the past decade \citep{cefis2022football}. Various metrics and methods are well established in soccer performance analysis, designed to capture the complexities of game dynamics. One of the most widely adopted models in modern soccer analytics is Expected Goals, commonly referred to as \textit{xG} \citep{fbref, cefis2025new}. This model estimates the probability that a given shot will result in a goal by considering factors such as shot distance, location, angle, whether it is a header, and the number of defenders between the attacker and the goalkeeper, among others. By quantifying the quality of scoring chances with numerical values, \textit{xG} provides a method for assessing how effectively teams create goal-scoring opportunities. Beyond shot characteristics, some models also incorporate variables that serve as proxies for psychological influences, such as match attendance, game importance (based on tournament stakes), and goal differential \citep{expectedGoal}.

With regard to similar metrics in other sports, \citet{macdonald2012expected} introduced a model to calculate expected goals in hockey based on statistics such as faceoffs, hits, and shots, facilitating performance evaluation at both the team and player levels (using adjusted plus-minus for expected goals). In rugby, \citet{kempton2016expected} derived an expected point value for a possession by considering contextual factors such as field location and the outcome of the preceding possession. In American football, \citet{yurko2019nflwar} developed a model to calculate added expected points—and even \textit{added win probability}—for each play, accounting for game context such as down and distance, field position, and remaining game time, among other factors.

Although extensive research exists on Expected Goals and analogous models—such as Expected Pass (\textit{xPass}) \citep{StatsBomb_ExpectedPass}, Expected Threat (\textit{xT}) \citep{van2020valuing}, and other possession value models \citep{StatsBomb_PossessionValue}—the soccer analytics literature remains relatively sparse when it comes to statistical adjustments of existing metrics for game context. The closest example to this work is , where offensive statistics accumulated over time intervals during specific game contexts were similarly projected onto a baseline scenario. While the core underlying idea is the very similar, our approach employs a far more granular, minute-by-minute game event dataset, allowing us to capture potential shifts in game dynamics between earlier and later stages of a match—something \citet{cemek2025statistical} does not address, as it omits the timing of events. Our modeling framework is also considerably more robust and flexible, leveraging Generalized Additive Models instead of basic dummy-variable encoding for scoring and red card contexts. Among other works involving statistical adjustments in soccer, a notable example is the use of bivariate Weibull distributions to model the relationship between goals scored and conceded, as seen in \citet{boshnakov2017bivariate}. Although these models are applied to soccer, they still often draw from frameworks developed in other sports—such as the "offense-defense" model introduced by \citet{harville1977use} in American football, where a team’s scoring output is modeled based on both its offensive ability and the opponent’s defensive strength. Additionally, acknowledging the impact of home-field advantage \citep{edwards1979home}, these models typically include adjustments for whether a team is playing at home.

    \subsection{Current Research Question}

This paper aims to address the limitations of traditional soccer statistics such as total shot attempts and corner kicks. Although not as pivotal as goals or goal-related metrics—the primary focus of most soccer analytics research (e.g., \textit{xG}, \textit{xPass}, \textit{xThreat})—these complementary statistics can still help construct a narrative about which team was more in control of the match flow. When game context is ignored, total shots and corner kicks can be misinterpreted, leading to conclusions that a team won (or lost) primarily due to luck, despite generating fewer (or more, respectively) scoring opportunities. The primary objective of this work is to develop a statistical adjustment that normalizes team performances in these complementary categories to a common baseline scenario, i.e., a tied home game with an equal number of players. This adjustment enables fairer comparative analysis and reduces the likelihood of such misinterpretations.

Beyond the adjustment itself, a key contribution of this work is a comprehensive investigation into how game context affects statistical categories such as shots and corners across five major European leagues over the past 15 years. Specifically, we hypothesize that certain tactical decisions are influenced by factors including score differential, red card differential, home versus away status, relative team strength (derived from prematch betting odds), and game minute. As noted earlier, a similar analysis was conducted in \citet{cemek2025statistical}, but it employed a much simpler model—using dummy-variable encoding rather than Generalized Additive Models (GAMs)—and focused solely on cumulative statistics over time spent in various game states. Notably, \citet{cemek2025statistical} did not account for the game minute at which events occurred, thereby failing to distinguish between actions taken early in the match and those occurring toward the end.

By leveraging detailed minute-by-minute match event data, our analysis accounts for both the half of play and the precise game minute. In addition to capturing broad trends in how offensive output evolves over the course of a match, this framework allows us to test specific hypotheses about interactions between contextual factors. For instance, it is intuitive to assume that trailing teams become increasingly aggressive in order to catch up in score, while leading teams would potentially grow more conservative in order to protect their lead. With our data and modeling approach, we can formally test whether such interaction effects are supported by empirical evidence. Given that the task of modeling minute-by-minute offensive output is more complex than analyzing aggregate counts over broader time intervals, we made sure to evaluate several candidate models before selecting our final modeling approach. That included Poisson, Negative Binomial, Zero-Inflated Poisson — potentially relevant due to the high frequency of single minutes with zero shot attempts or corner kicks — Gaussian, and Log-Link Gaussian.

Lastly, whereas much of soccer analytics relies on complex and less interpretable machine learning algorithms, we prioritize clarity by explicitly explaining \textit{how} various contextual factors affect offensive statistics, ensuring a full understanding of their effects and testing for their statistical significance. The additivity assumption in Generalized Additive Models (GAMs) allows us to interpret the partial effect of each covariate consistently, with statistical inference methods applying naturally due to GAMs being grounded in the linear modeling framework. Moreover, GAMs offer a clear mechanism for conducting significance testing and visualizing interaction effects, which is particularly valuable in evaluating the plausibility of several key hypotheses.

\section{Methods}

    \subsection{Data Collection and Cleaning}
    \label{sec:DataCollectionCleaning}

We compiled a large dataset by web scraping minute-by-minute event commentary from \textit{ESPN.com}, covering 15 seasons (2008-2023) in five major European football leagues: England's Premier League, Spain's La Liga, Germany's Bundesliga, France's Ligue 1, and Italy's Serie A. The events captured included various game statistics such as goals, shot attempts, corner kicks, and disciplinary actions (e.g., yellow and red cards), along with the half of play and the exact game minute in which each event occurred. In parallel, betting coefficients for the same leagues and period were collected from \textit{Oddsportal.com}. These coefficients serve as a reliable indicator of team strength, reflecting a combination of historical performance, recent form, injuries, expert assessments, and market trends, often being highly predictive of the game outcome \citep{lopez2018often}. To handle the dynamic nature of \textit{Oddsportal.com}, we employed the \textit{Selenium} browser automation tool \citep{noauthor_selenium_nodate}. This additional data was instrumental in accounting for potential confounding effects related to team-level characteristics.

Given the unstructured nature of the web-scraped text data, we undertook an extensive data wrangling and preprocessing pipeline. This included using regular expressions to identify and link game events to their corresponding teams, synchronizing the text commentary from \textit{ESPN.com} with the betting data from \textit{Oddsportal.com}, and addressing data inconsistencies. For instance, some games contained commentary from a different match or had inconsistent data formats (e.g., variations in parentheses usage and nomenclature). These issues affected only about 70 games out of over 25,000, and we opted to exclude them from the analysis. Additionally, we encountered occasional instances of out-of-order game minutes (e.g., minute 50 followed by minute 47 or a "Game ends" entry at the start of a half), which were resolved without removing the affected observations.

\begin{figure}
\centering
    \includegraphics[scale=0.175]{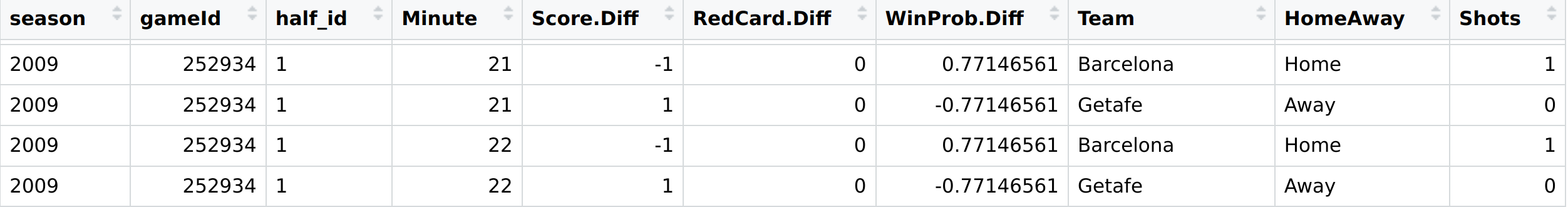}
    \caption{Minute-by-minute event data format for statistical modeling, with shot attempts as the response variable.}
    \label{fig:Data_Format}
\end{figure}

For statistical modeling, we structured the final dataset as shown in Figure \ref{fig:Data_Format} by aggregating the total events of each type (e.g., shot attempts, corner kicks) per minute per game, while also capturing relevant contextual factors such as homefield indicator, score and red card differentials at each minute. Additionally, we converted prematch betting coefficients into win probabilities and calculated the win probability differential between the teams to measure their relative strengths. % The primary objective was to assess each team's minute-by-minute offensive output rates (shots and corner kicks) while accounting for variables such as score and red card differential, home advantage, win probability differential, and game minute. 
The game minute resets to 1 at the start of the second half, with both the half indicator and the minute value used to represent the current game minute during modeling.

In total, we ended up with: 868,150  minute-by-minute observations (4,540 games) from  German Bundesliga, 1,080,842 (5,593 games) observations for Italian Serie A, 1,066,160 observations for Spanish La Liga (5,534 games), 1,042,564 observations (5,409 games) in French Ligue 1 and 918,658 observations (4,676 games) in English Premier League.

\subsection{Modeling Setup}
\label{sec:ModelSetup}

To model count-based statistics such as shot attempts and corner kicks, we primarily considered the Poisson distribution as the foundational approach. Specifically, in addition to regular Poisson regression, we explored the following alternatives: Negative Binomial distribution, which handles overdispersion (when the variance of the response variable exceeds its mean, violating one of the key assumptions of the Poisson distribution); and the Zero-Inflated Poisson approach, due to the high prevalence of zero counts in our response variables (over 85\%) \citep{mccullagh2019generalized}. We also considered Gaussian and Log-Link Gaussian regression models as a sanity check, to compare against the count-based models. For more details on model selection, see Section \ref{sec:ModSel}.

In our regression analysis, we included a baseline set of predictors to explain the response variable (e.g., shot attempts). These predictors were: score differential (\textit{Diff.S}), red card differential (\textit{Diff.RC}), home-field indicator (\textit{Home}), win probability differential (\textit{Diff.WP}), half indicator (\textit{Half}), and game minute (\textit{GM}). To model the dependence between observations for the same team and season, we incorporated team- and season-level random intercept effects. Additionally, to capture potential non-linear relationships, we used smoothing spline Generalized Additive Models (GAMs) \citep{wood2017generalized}. The model formula for shot attempts (\textit{Shot.Att}), which incorporates the baseline predictors along with the team- and season-level random effects, is as follows:

\begin{multline}
    \log(Shots) \sim s(Diff.S) + s(Diff.RC) + Home + s(Diff.WP) + s(GM \ | \ Half=1) + s(GM \ | \ Half=2) + \\
     s(Team) + s(season),
    \label{eq:ModelFormula}
\end{multline}

where the $s()$ notation represents smoothing splines used to model the non-linear effects of all predictors except the binary home-field indicator, which is represented by a dummy variable. To account for differences in game dynamics between the two halves, we model the game minute effect for each half separately using $s(GM \ | \ Half = 1)$ and $s(GM \ | \ Half = 2)$ terms. Additionally, the team- and season-level random effects are modeled via smooth terms that allow varying intercept values across different levels of the respective categorical variables (for more details, see \citep{wood2017generalized}). This combination of techniques enabled us to effectively model offensive production rates per minute while accounting for various contextual factors, as well as team- and season-level dependence in the data. 

Note that we also conducted significance testing for within-game dependence by introducing a game-level random effect (based on the \textit{gameId} variable from Figure \ref{fig:Data_Format}) instead of the season-level effect in (\ref{eq:ModelFormula}). We performed these tests for each individual season, swapping the $s(season)$ term for $s(gameId)$, but found that the game-level random effects were not statistically significant\footnote{Here, and throughout the manuscript (unless explicitly specified otherwise), the significance level is set to $0.05$.} for the vast majority of the $5 \times 15 = 75$ tests (five leagues, 15 seasons each). Therefore, we proceeded with only team- and season-level random effects to account for observation dependence.

\subsection{Model Selection}
\label{sec:ModSel}

For goodness-of-fit tests and model diagnostics, we used the simulation-based residuals approach \textit{DHARMa} \citep{hartig2017package}, which emulates the residuals-vs-fitted and quantile-quantile diagnostics from classic regression, but for a broader set of response variable distributions, including count-based models. In addition to diagnostic plots, \textit{DHARMa} provides significance tests for residual uniformity (Kolmogorov-Smirnov), overdispersion, and outliers. For Poisson-based models, it also includes an explicit test for zero-inflation. We used all of these tools to assess the appropriateness of the distributional classes considered for the data.

Among other model comparison tools, we used both the Akaike Information Criterion (AIC) and Bayesian Information Criterion (BIC) to prioritize simplicity and interpretability while maintaining a good fit \citep{akaike1998information, schwarz1978estimating}. Additionally, to explicitly evaluate out-of-sample performance, we adopted a leave-one-season-out cross-validation (LOSOCV) approach, inspired by \citet{yurko2019nflwar}. The evaluation metrics included mean absolute error (MAE) and root mean squared error (RMSE) on the response scale, which represent the average errors when predicting shot attempts and corner kicks for each game minute.

To select the degrees of freedom for the smoothing spline estimates, we used generalized cross-validation (GCV), the default method in the \textit{mgcv} package. This method efficiently calculates a leave-one-out cross-validation smoothing spline estimate with the same computational cost as obtaining a single fit \citep{wood2017generalized}. After observing instability in the smoothing spline fits at the extreme ends of certain predictor variable ranges, we considered several ad-hoc approaches to address this issue. First, similar to \citet{cemek2025statistical}, we "binned" extreme values of predictors into a single value, in order to to treat them the same during modeling. For example, score differentials of $3$ and higher ($3,4,5,\dots$) were treated as $3$, while score differentials of $-3$ and lower ($-3,-4,-5,\dots$) were treated as $-3$. Score differentials of $-2$, $-1$, $0$, $1$, and $2$ remained unchanged. For red card differentials, the "$+2$" and "$-2$" categories were binned, and for game minutes, extra time (i.e., minutes past the 45-mark) was treated the same as minute 45. The intuition was to alleviate fitting issues primarily caused by lower and/or inconsistent sample sizes at the extreme ends of the ranges. 

Additionally, as a lower-complexity alternative to the smoothing spline fit, we considered an ad-hoc natural cubic spline approach \citep{wood2017generalized}. Specifically, we manually set the knots to the most prevalent discrete values of score differential ($-2, -1, 0, 1, 2$) and red card differential ($-1, 0, 1$). For the effect of game minute, we chose a slightly lower degree of freedom compared to the smoothing spline estimate (e.g., 5 as opposed to 8), and we made the effect of win probability differential fully linear, as it appeared linear in the smoothing spline estimates. Finally, we compared all models (smoothing splines GAM with/without binned extremes, natural cubic splines with/without binned extremes) using AIC, BIC, and LOSOCV.

Lastly, to compare and illustrate how well-calibrated our models were in predicting event rates across the five soccer leagues, we adapted the "leave-one-season-out" approach presented by \citet{yurko2019nflwar}. The main difference is that we applied this approach to count data, whereas Yurko et al. used it for ordinal regression modeling to predict rates of various ordinal outcomes. Specifically, we bucketed the predicted event rates into intervals of size 0.01 (i.e., [0, 0.01], (0.01, 0.02], etc.) and plotted the midpoints of these intervals against the actual event rates observed for corresponding data points. For a well-calibrated model, the predicted and actual event rates should form an $x = y$ diagonal line.

\subsection{Variable Selection and Hypothesis Testing}
\label{sec:VarSelHypothesisTest}

To evaluate the importance of all variables under consideration, we conducted significance testing for each factor in the baseline set: score differential, red card differential, home advantage, win probability differential, and game minute. Given the large sample sizes for each league ($\sim$ 1,000,000 observations), even small effects are more likely to be statistically significant. To address this, we fit the models separately for each season, removing the season-level random effect from (\ref{eq:ModelFormula}) for this task, resulting in approximately 50,000–75,000 observations per fit. We also applied Holm's multiple-testing correction \citep{holm1979simple} to control the Type I error rate across all $5 \times 15 = 75$ tests (five leagues, 15 seasons) for each respective effect.

Additionally, we tested the significance of pairwise factor interactions by including each interaction as a separate term alongside the baseline set of five key factors for each season, and then evaluating the improvement achieved. This approach allowed us, among other research questions, to investigate hypotheses about how the effects of score and/or red card differential on offensive output change during different stages of the game. We then compared and contrasted the effect displays for cases where interactions were statistically significant versus those where they were not.

\subsection{Statistical Adjustment Approach}
\label{sec:StatAdj}

After fitting the final model and obtaining estimates for the effects of score differential, red card differential, and home advantage, we proceeded to develop an adjustment that projects offensive outputs onto the shared scenario of a tied home game played at even strength throughout. Suppose the offensive output, such as shot counts, is accumulated during a period when the score differential is $s$, with $s \neq 0$. In this case, we adjust the shot count for a tied game scenario via multiplying it by $e^{-(\hat{\eta}_s - \hat{\eta}_0)}$, where $\hat{\eta}_s - \hat{\eta}_0$ represents the difference in the estimated linear predictor values ($\eta = \log(\text{Shots})$) between a score differential of $s$ and $0$. This adjustment captures the inverse effect of transitioning from a tied game to a differential of $s$.

Similarly, if the shot count is recorded during a period with a red card differential $rc$, $rc \neq 0$, we project it onto a scenario where the red card differential is zero by applying the same type of reciprocal adjustment. Notably, since the final model ends up not including interaction terms (see discussion and reasoning later), the linear predictor estimate differences for various score and red card differentials remain consistent as long as all other predictors are held at the same values for both scenarios. Lastly, to adjust an away team’s performance to a hypothetical home game, we multiply their shot counts by $e^{-\hat{\gamma}}$, where $\hat{\gamma}$ is the estimated coefficient for the home dummy variable, with home serving as the baseline category. If a team's offensive output was accumulated in a tied home game with no red card differential, no statistical adjustment is required.

In addition to the point estimates for projected values, we also provide $95\%$ confidence intervals (CIs). These intervals are initially calculated for the difference in linear predictor values, such as $\hat{\eta}_s - \hat{\eta}_0$ for the score differential $d$, where $d \neq 0$. We utilize the Bayesian posterior covariance matrix for the GAM model parameters to estimate the standard error \citep{wood2017generalized}. Subsequently, we apply the appropriate quantiles from the standard normal distribution to obtain the $95\%$ confidence interval for the difference, which is then transformed to the scale of the response.

Note that no explicit statistical adjustment is made for win probability differential, half of play, or game minute. We believe it would be unfair to penalize a team for being inherently stronger (or reward a team for being inherently weaker), and it does not make practical sense to adjust statistics solely based on when they were accumulated during the match. The primary purpose of including these variables in the model was twofold: first, to control for them when estimating the effects of other factors, such as score differential, red card differential, and home advantage; and second, to test for potential interactions with these factors, which could affect how score or red card differential impacts offensive output at different values of win probability differential or game minute.

To illustrate the intuition behind the adjustment mechanism, we showcased the strongest value shifts in shots and corner kicks resulting from our adjustment—both within individual games and across an entire season. We accompany these shifts with relevant contextual information, such as statistics accumulated while a team was ahead or behind in score, up or down in men, and whether they played at home or away. In addition to highlighting the strongest individual shifts, we conducted a correlation analysis to demonstrate that the adjusted values align more closely with intuitive statistical categories, including final league standings in terms of points earned (a team earns 3 points for a win, 1 for a draw, and 0 for a loss). For each season, we calculated the correlation between each team’s actual values (shots or corner kicks) and the points earned, repeating the same calculation for the adjusted values. We then averaged these correlations across all 15 seasons and all five leagues under consideration, reporting the overall mean correlation ($\text{mean}(\text{Err})$) along with the standard error estimate ($\frac{\text{sd}(\text{Err})}{\sqrt{n}},$ where $n = 15 \times 5 = 75$).

Lastly, as another way to demonstrate that our adjustments improve upon the raw statistics, we evaluated their performance in forecasting score differentials using a basic multiple linear regression model. Specifically, the model included the home team's and away team's statistic values (e.g., shots per game up to that point in the season) as explanatory variables, with the response variable being the final score differential of the game from the home team's perspective. For each season, we used the first 50\% of the games (approximately 190 games total, or about 19 games per team) to train the model, and the remaining 50\% to test forecasting performance. The per-game averages used for prediction were calculated solely from the training data to prevent data leakage. We used root mean squared error (RMSE) as the performance metric, comparing it between regression model that used adjusted statistic values versus the one using actual values. For each league, we conducted a paired $t$-test on the differences in RMSE between the adjusted and actual statistics for each year, testing whether the mean difference was significantly different from zero. To verify the normality assumption, we applied the Shapiro-Wilk normality test \citep{shapiro1965analysis}, and we report the corresponding $p$-values and 95\% confidence intervals.

\section{Results and Discussion}

\subsection{Model Selection}
\label{sec:ModSel}

Using the modeling formula 
(\ref{eq:ModelFormula}) outlined in Section \ref{sec:ModelSetup}, we fitted Poisson, Negative Binomial, Gaussian, and Log-link Gaussian GAMs to each season from 2008 to 2023 for the five major European soccer leagues under consideration, and conducted DHARMa quality-of-fit tests and diagnostics. We found the count-based models to be more appropriate than their Gaussian counterparts, as the latter consistently showed statistically significant deviations in the Kolmogorov-Smirnov test and displayed inadequate patterns in both residuals-vs-fitted and quantile-quantile plots. In comparing the Poisson and Negative Binomial fits, both models consistently satisfied the Kolmogorov-Smirnov test. However, the Poisson model exhibited statistically significant violations in both overdispersion and zero-inflation tests, which the Negative Binomial model did not. See the supplementary materials for Holm-corrected p-values of all the aforementioned goodness-of-fit tests, along with several examples of DHARMa diagnostic plots and their comparisons across the models.

\begin{figure}[]
\centering
        \includegraphics[scale=0.95]{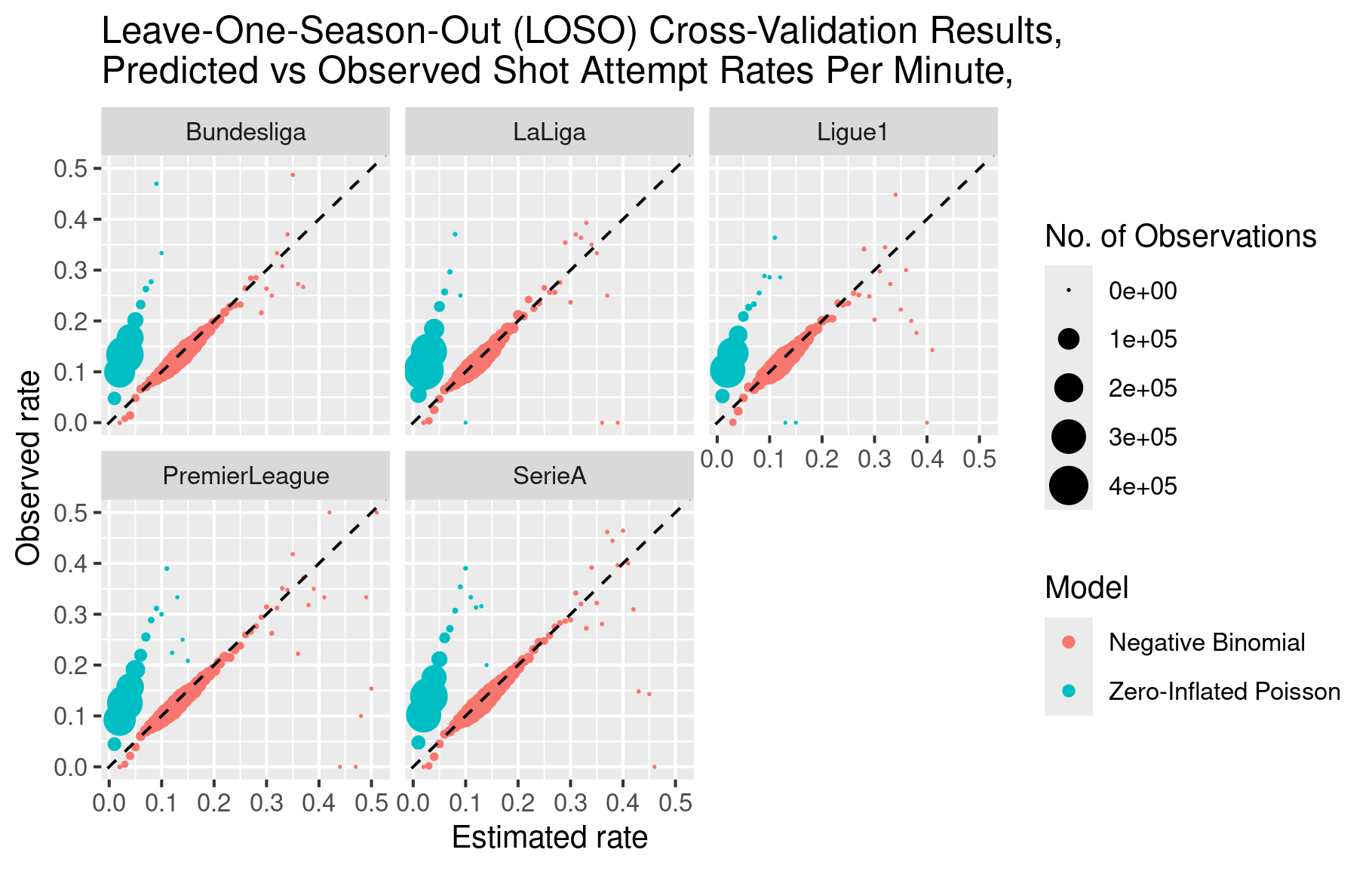}
\caption{Leave-one-season-out cross-validation calibration results for the final Negative Binomial regression model when predicting shot attempts across five major European soccer leagues and 15 seasons under consideration. Zero-Inflated Poisson regression model is provided as well for comparison.}
\label{fig:Calibration_Plots}
\end{figure}

In comparing the Negative Binomial and Zero-Inflated Poisson (ZIP) regression models, due to lack of DHARMa diagnostic tools for Zero-Inflated Poisson GAM model, we resorted to AIC, BIC and LOSO calibration approach described in Section \ref{sec:ModelSetup}. As shown in Tables \ref{tab:AIC_Pois_NegBin_ZIP} and \ref{tab:BIC_Pois_NegBin_ZIP} in the Appendix, AIC indicated a slight advantage for ZIP, while BIC favored the Negative Binomial, mostly due to the significantly lower complexity of the Negative Binomial model (ZIP has twice as many parameters due to its two components). The LOSO calibration approach, illustrated in Figure \ref{fig:Calibration_Plots}, demonstrated that, except for a few cases with a really low number of observations, the Negative Binomial's out-of-sample event rate predictions closely aligned with the actual observed rates, in contrast to the ZIP model (the plot is for shot attempts, but the same pattern holds for corner kicks). This discrepancy largely results from ZIP overly emphasizing individual zero counts through its zero-inflation component, while underestimating actual event rates by multiplying the Poisson count probability by the probability of not having a structural zero. Since one of the main priorities of our statistical adjustment mechanism is to accurately reflect the \textit{rates} at which events occur in various game contexts, also combined with the fact that BIC clearly favored the Negative Binomial model for its lower complexity while maintaining a good fit, we chose Negative Binomial as the response distribution moving forward.

After fitting a single Negative Binomial smoothing spline GAM to all 15 seasons for each respective league, we observed that effect estimates were quite unstable at the extreme ends of the predictor range (see Figure \ref{fig:EffectDisplays} in the Appendix). For example, the effect of score differential exhibited an inconsistent pattern starting at $\pm 4$, while the effect of game minute beyond 45 appeared completely unpredictable. These irregularities were likely a byproduct of the leave-one-out cross-validation estimates used by default when fitting smoothing spline GAMs. While computationally efficient, these estimates tend to be more prone to overfitting. To address this, we considered ad-hoc approaches such as binning extreme categories and creating a simpler natural cubic spline model (for more details, see \ref{sec:ModSel}). After evaluating both approaches using AIC, BIC, and LOSOCV, the binning approach with smoothing spline GAMs performed best, and we selected it as our primary model moving forward. For more details on the comparison between the considered models and the potential impact on the nature of the effects, see the supplementary materials.

For the final Negative Binomial GAM with binned extreme categories, the leave-one-season-out predictive performance of the model was as follows. When predicting per-game shot counts (the sum of single-minute predictions across the entire game), the average errors across leagues ranged from 3.24 to 3.53 shots per game, which corresponds to about 27-28\% of the actual per-game shot attempt averages. For corner kick predictions, the average errors across leagues were between 1.9 and 2.1 corners per game, about 40\% of the actual per-game corner kick averages. The out-of-sample $R^2$ values for shot attempts ranged from 21\% to 31\%, while for corner kicks, they ranged from 17\% to 22\%, depending on the league. Overall, these performance metrics are not necessarily impressive, but it's important to note that the set of features used in our model is relatively limited (lacking fine-grained data such as player rosters, injuries, weather factors, etc.), and the additive modeling framework is simpler than some other existing alternatives. This simplicity was intentional, as we prioritized interpretability and intuitiveness in the adjustment mechanism. While pure out-of-sample predictive performance could likely have been improved by considering other model types (e.g., random forests, neural networks), these "black box" models would have made the effects harder to interpret, and the adjustment mechanism would have been less intuitive (compared to the straightforward multiplication by a coefficient used in our case). Lastly, it's evident that corner kicks appear to be more challenging to predict than shot attempts.

\subsection{Nature of Estimated Effects}
\label{sec: NatureEffects}

\begin{figure}[]
    \centering
   \includegraphics[scale=0.235]{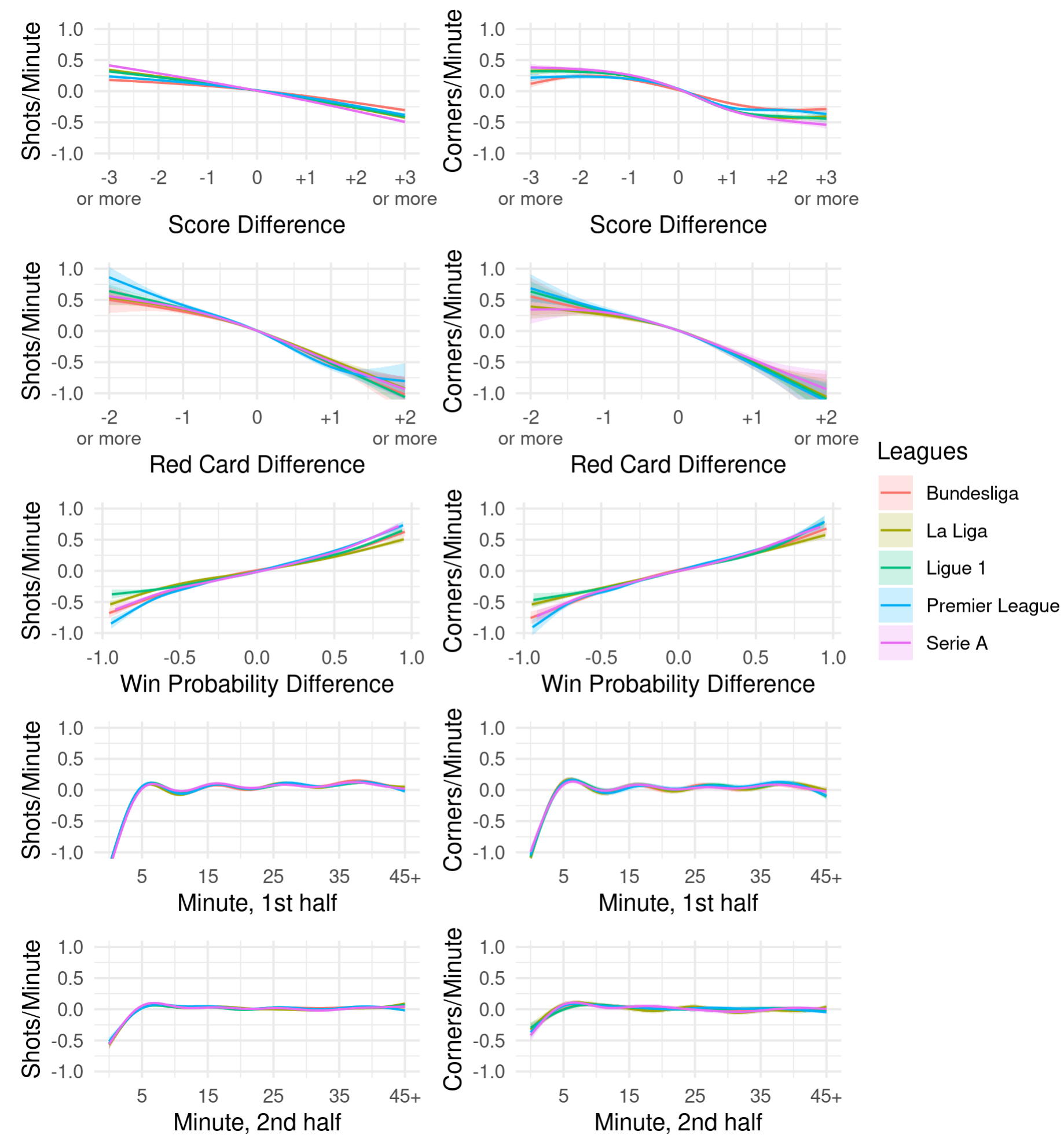}
    \caption{Nature of effects for score differential, red card differential, win probability differential and game minute in 1st/2nd halves on shot attempts (left) and corner kicks (right) in the baseline Negative Binomial Generalized Additive smoothing splines model fitted to 15 seasons (2008-2023) across five major European soccer leagues. The y-axis is on the scale of linear predictor (log-response). \textbf{The extreme predictor values ($\pm3$ for score differential, $\pm2$ for red card differential, $45+$ for minute) were binned}.}
    \label{fig:EffectDisplays_BINNED}
\end{figure}

Figure \ref{fig:EffectDisplays_BINNED} illustrates the effects of the four baseline factors modeled with smoothing splines (all except the home-field indicator, which was modeled using a dummy variable), with separate smooths for two halves of play in case of game minute.

 After binning the extreme categories, score differential exhibits a clear negative pattern, where offensive production tends to decrease as the score difference becomes more positive. The red card differential shows a strong negative effect on offensive production, while the win probability differential has a strong positive effect, with both relationships appearing mostly linear. The effect of game minute shows offensive output gradually increasing over the first five minutes of each half, with lower activity at the start of the 1st half compared to the 2nd. This makes sense, as teams may use the first few minutes of the game to assess each other, while in the 2nd half, they are more calibrated for the opponent. After the first five minutes, there is a relatively stable, flat pattern with the same level of activity for the remainder of either half.

One curious observation is that the Italian Serie A appears to have the steepest impact of score differential on offensive production. Leading teams seem to play even more defensively than in other leagues, limiting their own offensive production while also allowing their opponents to accumulate even more offensive output (this is also reflected in their multiplicative coefficients in Figure \ref{fig:MultCoef_BINNED} later). Historically, Italian soccer has been strongly associated with defensively-minded tactics, the most famous being "catenaccio" — a system that has been embraced by the Italian league and national team throughout their history \citep{sosa2015identidad}. While catenaccio in its original form is less prevalent in recent decades \citep{trequattrini2016does}, modified versions of it, with an overall emphasis on defensive responsibility, have likely remained in Italian soccer. We suspect that these plots may be reflecting this tendency.

\subsection{Studying Interaction Effect Hypotheses}

\begin{figure}[h]
   \centering
    \includegraphics[scale=0.85]{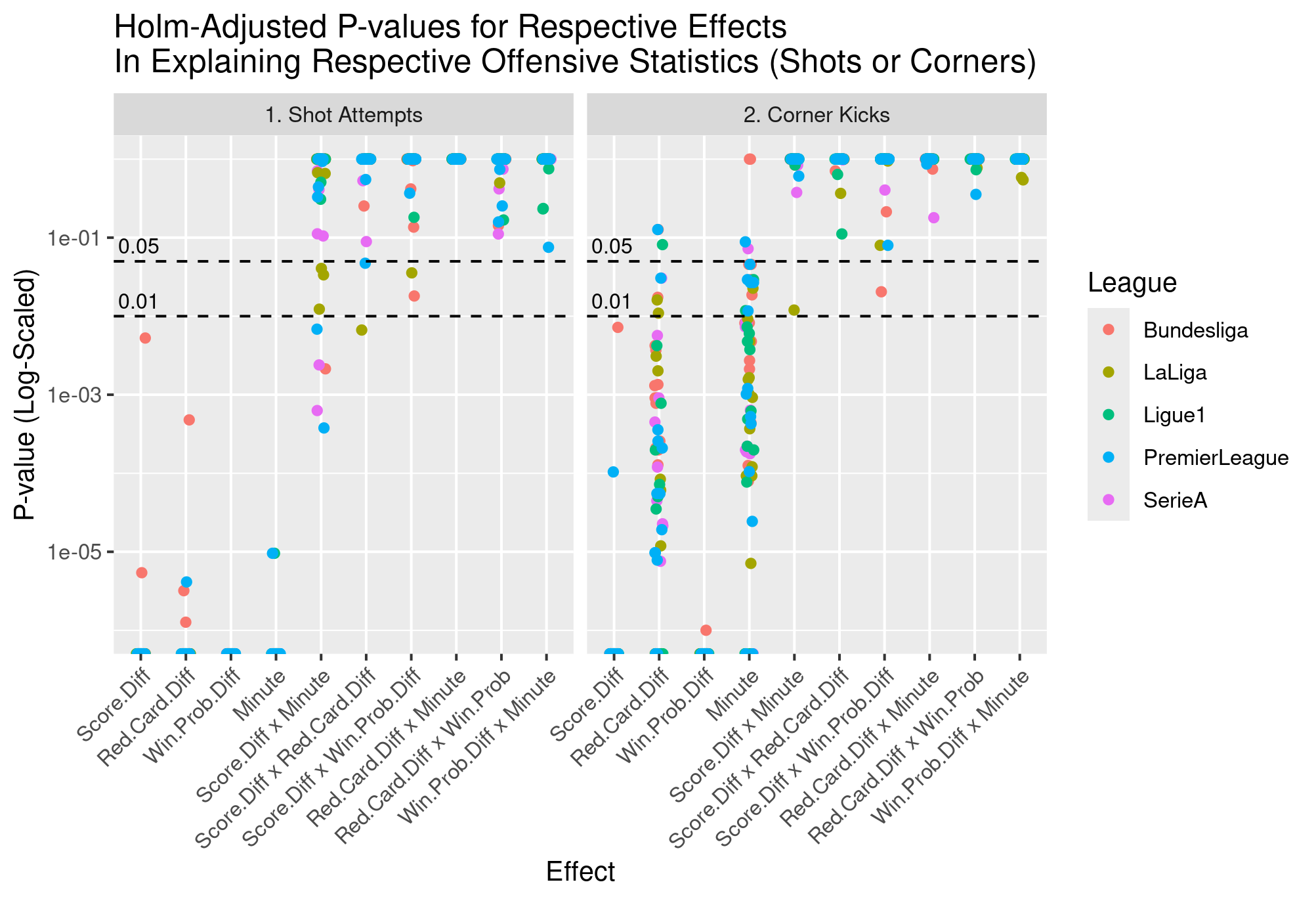}
   
    \caption{The p-values from season-by-season fits of Negative Binomial smoothing splines GAM model with binned extreme values. Each effect's p-value was Holm-adjusted for the fact that there were $15\times 5=75$ tests being conducted. Five key factors were tested in a model without the interactions included. Each interaction was tested as an additional term to the baseline model of five key factors.
    }
    
    \label{fig:PvaluesHypothTestingFiveKeyFactorsAndInteractions}
\end{figure}

After performing season-by-season hypothesis testing to assess the statistical significance of each of the five key factors, along with their pairwise interactions, the results are shown in Figure \ref{fig:PvaluesHypothTestingFiveKeyFactorsAndInteractions}. Accounting for Holm's correction, every individual key factor under consideration (score differential, red card differential, game minute, win probability differential, home/away factor) was always statistically significant when modeling shot attempts, and significant in the vast majority of cases for corner kicks. This confirms our intuition to retain all of these factors in our baseline model.

\begin{figure}[]
   \centering
    \includegraphics[scale=0.8]{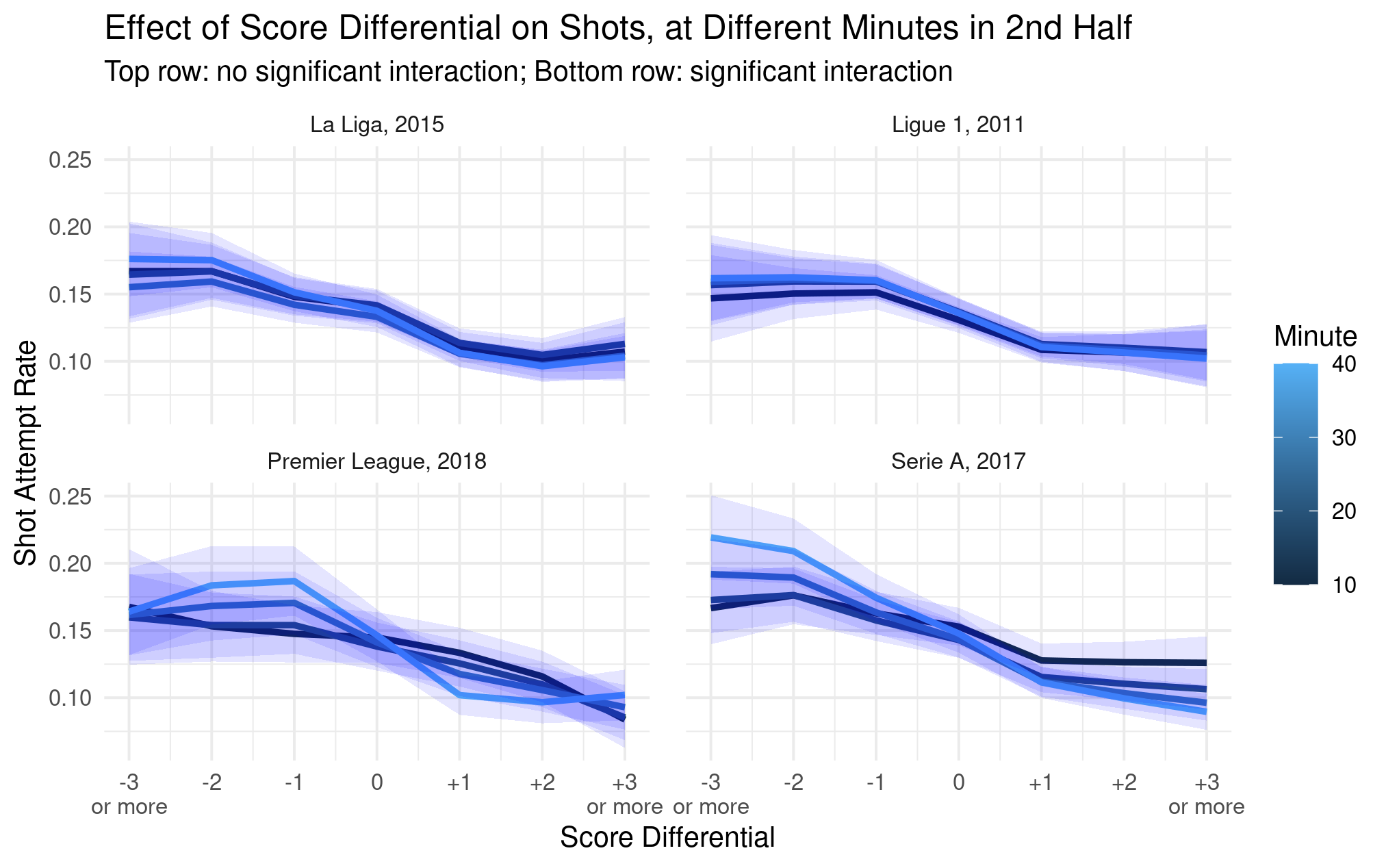}
   
    \caption{Investigating the nature of the potential interaction effect between score differential and game minute.
    }
    
    \label{fig:InteractionEffect}
\end{figure}

 As for the pairwise interaction terms, there were very few cases (virtually none for corner kicks) where adding an interaction term resulted in a statistically significant improvement over the baseline model. In particular, the only interaction term with at least a handful of significant tests was the one between score differential and game minute when predicting shot attempts, which partially confirms one of our primary intuitive hypotheses. To investigate the nature of the effect and confirm that it aligns with our intuition, Figure \ref{fig:InteractionEffect} illustrates the effects of score differential on shot attempts at different points in the game, for a couple of leagues and years where the interaction was found insignificant (top row) and significant (bottom row), respectively. For the cases where it was significant, it indeed confirms our intuition: the trailing team becomes more active toward the latter stages of the second half (end of the game), while the leading team becomes more conservative. This is particularly true for smaller score differentials, like $\pm1$, where the game is still within reach. However, for the leagues and years where the interaction between score differential and game minute was not significant— which is the vast majority—it is clear that the data did not support the effect, indicating that a simpler model with additive effects was sufficient. Therefore, for consistency in our adjustment mechanism, we proceeded with the baseline model of the five key factors moving forward.

Meanwhile, for corner kicks, there were even fewer cases where any of the interactions were found significant after Holm's correction. More notably, in some cases, even the additive effects of red card differential and game minute had p-values above the typical thresholds of 0.01/0.05, which was never the case for shot attempts. This indicates that while corner kicks mostly exhibit similar patterns to shot attempts, they are not as straightforward to model as a function of those game context variables. This manifests in lower quality-of-fit and out-of-sample performance metrics, as discussed at the end of Section \ref{sec:ModSel}.

\subsection{Statistical Adjustment}

\begin{figure}[h]
\includegraphics[scale=0.9]{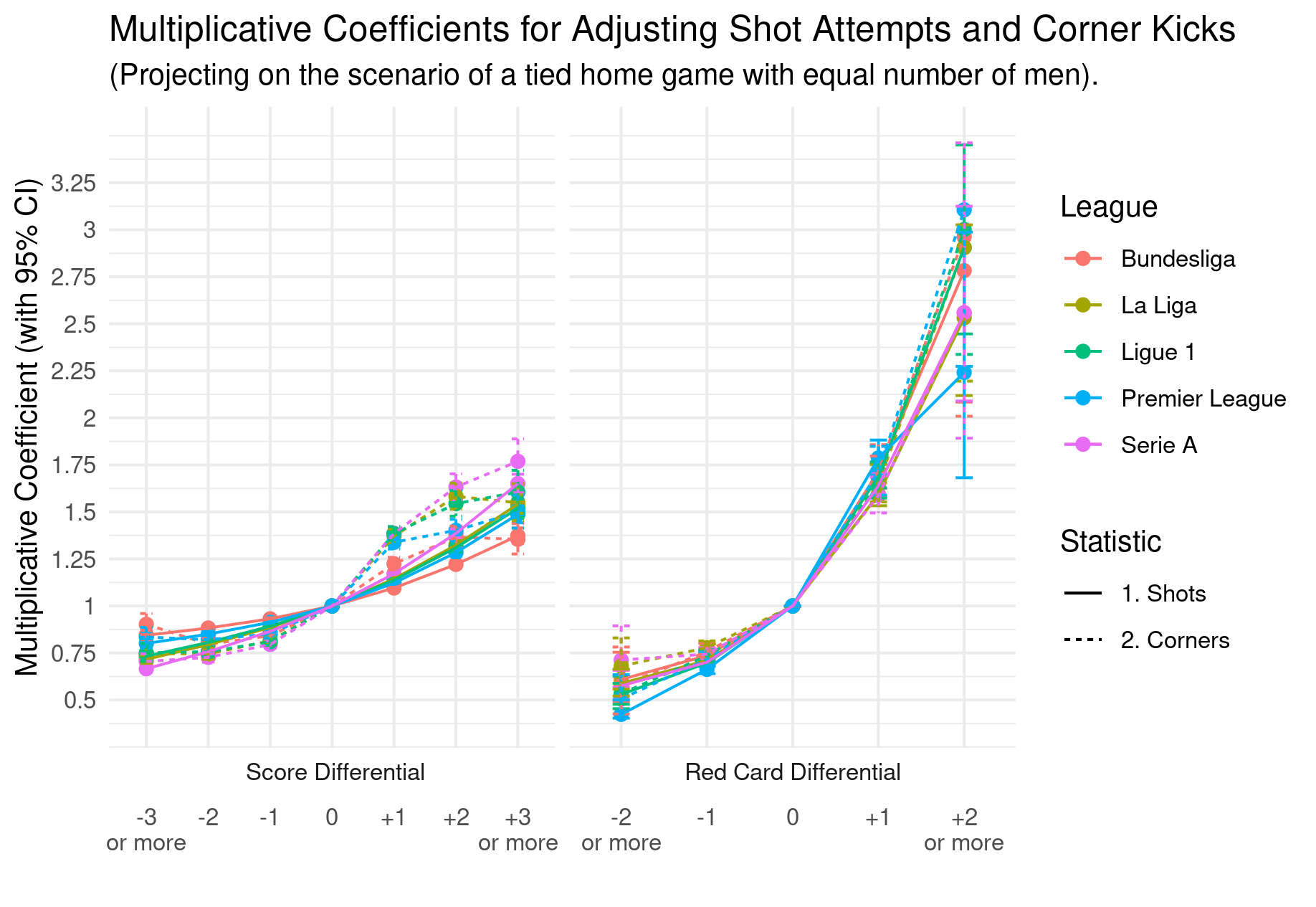}
   
    \caption{Multiplicative coefficients for shot attempts and corner kicks generated during respective score and red card differentials to project those on the scenario of a tied game with equal number of men, along with 95\% confidence intervals.}
    
    \label{fig:MultCoef_BINNED}
\end{figure}

Figure \ref{fig:MultCoef_BINNED} shows the multiplicative coefficients for shot attempts and corner kicks, derived by converting the effect estimates from Figure \ref{fig:EffectDisplays_BINNED} as described in Section \ref{sec:StatAdj}. Shots and corner kicks attempted while trailing receive less credit, with multiplication coefficients ranging from 0.70 to 0.90 (a reduction of 10-30\%). In contrast, offensive output produced when in the lead is weighted more heavily, with coefficients ranging from 1.10 to 1.75 (an increase of 10-75\%), depending on the league and size of the lead. Offensive output generated with a numerical advantage (negative red card differential) is significantly downgraded, with a 25\% reduction when up by one player (multiplication coefficient of 0.75), and a 30-60\% reduction when up by at least two players (coefficients of 0.40-0.70). Conversely, shots and corner kicks attempted with fewer players on the field receive substantial extra credit, with coefficients of 1.7-1.9 when down by one player, and 2-3 when down by at least two players.

These patterns are generally consistent across leagues, though the magnitudes vary, with Serie A showing steeper coefficients for score differential adjustments (see Section \ref{sec: NatureEffects} for the 'catenaccio' discussion). The wider confidence intervals for teams with at least a 2-man disadvantage likely result from smaller sample sizes and more inconsistent offensive activity among these teams, compared to those with a 2+ man advantage. In the following subsections, we examine the impact of these adjustments at both the single-game and season-long levels, highlighting the most significant shifts in values and the underlying intuition behind the adjustment mechanism.

\subsubsection{Single-Game Adjustments}

Tables \ref{tab:StatAdjustShots} and \ref{tab:StatAdjustCorners} present the individual games where a team experienced the most significant positive and negative adjustments to shot production and corner kicks, respectively, across the five European soccer leagues from 2008 to 2023. To clarify the adjustment mechanism, the tables also include a contextual breakdown of the game situations during which the respective offensive outputs were generated.

\begin{table}[]
\centering
\resizebox{\textwidth}{!}{
\begin{tabular}{|l|l|l|l|l|l|l|l|l|l|}
League & Team & Opponent & Season & \multicolumn{2}{c}{Total Shots} & \multicolumn{4}{c}{Actual Shots (\& Minutes Played) When} \\
 &  &  &  &  &  & Up & Down & Up & Down \\
 &  &  &  & Actual & Adjusted, [95\% CI] & \(1+\) goal & \(1+\) goal & \(1+\) man & \(1+\) man \\
\midrule
\multicolumn{3}{c}{\textbf{Strongest Positive Shifts in Each League:}} & & & & & & & \\
ENG\textsuperscript{1} & Manchester City & @ West Brom & 2012/13 & 24 & 41.7, [39.7, 43.9] & 1 (2) & 2 (13) & 0 (0) & 20 (78) \\
FRA & Caen & @ Troyes & 2015/16 & 15 & 31.7, [30.4, 33.1] & 13 (78)$^*$ & 0 (0) & 0 (0) & 9 (38) \\ 
GER & Bayern Mun & Stuttgart & 2020/21 & 15 & 31.9, [30.1, 33.7] & 12 (76)$^*$ & 0 (0) & 0 (0) & 14 (82)\\
ITA & Inter Milan & Chievo V & 2017/18 & 40 & 54.6, [53.6, 55.7] & 28 (71)$^*$ & 0 (0) & 0 (0) & 0 (0) \\
SPA & Real Madrid & @ Real Zaragoza & 2011/12 & 40 & 60.7, [59.1, 62.4] & 33 (67)$^*$ & 0 (0) & 0 (0) & 0 (0) \\
%... & ... &  ... & ... & ... & ... & ... & ... & ... & ... \\
\multicolumn{3}{c}{\textbf{Strongest Negative Shifts in Each League:}} & & & & & & & \\
ENG & Blackpool & W Brom & 2010/11 & 26 & 13.0, [11.1, 15.2] & 25 (79)$^*$ & 0 (0) & 26 (81)$^*$ & 0 (0)  \\
FRA & Bordeaux & Montpellier & 2021/22 & 31 & 13.8, [12.5, 15.1] & 0 (0) & 31 (94)$^*$ & 31 (66)$^*$ & 0 (0)  \\
GER & Eintracht & Stuttgart & 2010/11 & 30 & 22.1, [21.5, 22.8] & 0 (0) & 12$^*$ (26) & 26 (76) & 0 (0) \\
ITA & AS Roma & Venezia & 2021/22 & 43 & 28.7, [28.0, 29.4] & 0 (0) & 29 (79) & 38 (70) & 0 (0) \\
SPA & Getafe & Deportivo & 2009/10 & 22 & 12.8, [12.1, 13.6] & 0 (0) & 22 (79)$^*$ & 19 (66)$^*$ & 0 (0) \\
\end{tabular}}
\caption{Individual games across 2008-2023 seasons for each of the five European soccer leagues under consideration where a team experienced strongest positive (first five rows) and strongest negative (last five rows) absolute adjustments in their shot production. An "@" symbol indicates that the team played away, at the opponent's home field. An asterisk "*" denotes games where the score or red card differential exceeded \(\pm 1\) (e.g., ahead by at least 2 goals or down by at least 2 men).
}
\label{tab:StatAdjustShots}
\end{table}

\footnotetext[1]{ENG: English Premier League; FRA: French Ligue 1; GER: German Bundesliga; ITA: Italian Serie A; SPA: Spanish La Liga.}

\begin{table}[]
\centering
\resizebox{\textwidth}{!}{
\begin{tabular}{|l|l|l|l|l|l|l|l|l|l|}
League & Team & Opponent & Season & \multicolumn{2}{c} {Total Corners} & \multicolumn{4}{c}{Actual Corners (\& Minutes Played) When} \\
 & & & &  &  &  Up \ \ \ &  Down \  &  Up \ \ \ & \ Down \ \\
 & & & & Actual & Adjusted & $1+$ goal & $1+$ goal & $1+$ man & $1+$ man \\
\midrule
\multicolumn{3}{c}{\textbf{Strongest Positive Shifts in Each League:}} & & & & & & & \\
ENG & Man City & @ Norwich & 2012/13 & 9 & 17.2, [16.1, 18.4] & 9 (101)$^*$ & 0 (0) & 0 (0) & 4 (59) \\
FRA & Marseille & @ Guingamp & 2017/18 & 11 & 18.9, [17.9, 20.0] & 2 (53)$^*$ & 0 (11) & 0 (0) & 8 (31) \\
GER & VfL Wolfsburg & @ Eintracht & 2012/13 & 9 & 17.9, [16.5, 19.4] & 8 (87)$^*$ & 0 (0) & 0 (0) & 7 (62)\\
ITA & AS Roma & Palermo & 2012/13 & 15 & 25.6, [24.3, 26.9] & 13 (85)$^*$ & 0 (0) & 0 (0) & 2 (14) \\
SPA & Barcelona & @ Celta & 2020/21 & 8 & 22.3, [20.7, 24.0] & 8 (91)$^*$ & 0 (0) & 0 (0) & 7 (60) \\
% ... & ... &  ... & ... & ... & ... & ... & ... & ... & ... \\
\multicolumn{3}{c}{\textbf{Strongest Negative Shifts in Each League:}} & & & & & & & \\
ENG & Bournemouth & Tottenham & 2018/19 & 10 & 6.0, [5.2, 6.9] & 0 (5) & 0 (0) & 9 (58)$^*$ & 0 (0)\\
FRA & Paris St-Ger & St Rennais & 2012/13 & 18 &  9.5, [8.6, 10.5] & 0 (0) & 17 (71) & 16 (72)$^*$ & 0 (0)  \\   
GER & Eintracht & Stuttgart & 2010/11 & 13 &  9.1, [8.5, 9.6] & 0 (0) & 5 (26)$^*$ & 13 (76) & 0 (0) \\    
ITA & AS Roma & Venezia & 2021/22  &  20 & 13.1, [12.6, 13.8] & 0 (0) & 14 (79) & 18 (70) & 0 (0) \\
SPA & Getafe & Deportivo & 2009/10 & 14 & 8.7, [7.8, 9.6] & 0 (0) & 13 (79)$^*$ & 11 (66)$^*$ & 0 (0) \\
\end{tabular}
}
\caption{Individual games across 2008-2023 seasons for each of the five European soccer leagues under consideration where a team experienced strongest positive (first five rows) and strongest negative (last five rows) absolute adjustments in their corner kick production. An "@" symbol indicates that the team played away, at the opponent's home field. An asterisk "*" denotes games where the score or red card differential exceeded $\pm 1$ (e.g., ahead by at least 2 goals or down by at least 2 men).}
\label{tab:StatAdjustCorners}
\end{table}

The strongest positive adjustment in each respective league (top five rows) corresponds to the team that generated most of their offensive output while leading by at least one goal and having at least a one-man disadvantage. As shown previously in Figure \ref{fig:EffectDisplays_BINNED}, such game contexts (playing with a lead and with fewer men) typically hinder shot production, thus benefiting from an adjustment that projects them onto a more favorable scenario—a tied game at even strength. For example, in the German Bundesliga, the most significant increase in shot attempts occurred for Bayern Munich in their 4-0 win at home against Stuttgart during the 2020/21 season, where their shot count rose from 15 to 31.9 ($95\%$ CI: $[30.1, 33.7]$) due to the adjustment. The majority of their shots (14) were attempted after losing a player to a red card in the 12th minute, with 12 of those attempts coming after they took the lead, which they maintained. Both factors created unfavorable conditions for offensive production, making Bayern's performance even more remarkable. Regarding corner kicks, the largest increase in the Bundesliga occurred for VfL Wolfsburg in their 2-2 draw on the road against Eintracht. Wolfsburg led for the majority of the game while also receiving a red card in the first half. They generated 7 of their 9 corner kicks while playing both shorthanded and in the lead, doubling their count due to the adjustment, which acknowledged their impressive effort given the circumstances.

One commonality among the largest positive shifts is that the majority of these teams generated most of their offensive output after going up by at least two goals, which increases their weight in our adjustment approach. For example, in Table \ref{tab:StatAdjustShots}, both Bayern Munich and Caen had 11 of their 15 shots occur when leading by at least two goals, while for Inter Milan it was 26 of their 40 shots, and for Real Madrid — 31 out of 40. In Table \ref{tab:StatAdjustCorners}, AS Roma had 13 out of their 15 corner kicks when leading by at least two goals, while Barcelona had 6 out of 8. These projections suggest that the actual offensive outputs of these teams may underrepresent their performance level due to tactical considerations. They likely would have generated even more shot attempts if the game circumstances were more conducive to it, such as a tied game played at even strength.

%For the five strongest negative adjustments (last five rows of Tables \ref{tab:StatAdjustShots} and \ref{tab:StatAdjustCorners}), the teams 

For the strongest negative adjustment in each respective league (last five rows of Tables 1 and 2), the team accumulated most of their offensive output while trailing by at least one goal and playing with at least a one-man advantage. As shown earlier in Figure \ref{fig:EffectDisplays_BINNED}, both of these settings typically associate with increased shot production. Consequently, our adjustment lowered these shot counts by projecting them onto a less favorable scenario—a tied game at even strength. The most significant reduction in shots occurred for Bordeaux in their 0-2 loss at home against Montpellier during the 2021/22 Ligue 1 season. Their actual shot count of 31 was adjusted down to 13.8 ($95\%$ CI: $[12.5, 15.1]$). This was due to Bordeaux's 31 shots all occurring after they gained a two-man advantage (11 vs. 9, with Montpellier receiving a second red card by the 45th minute) while already trailing 0-2. In such a setting, high shot production is expected, even more so than with a one-man advantage or a one-goal lead. Thus, when projecting onto a tied game with equal men, our adjustment led to a substantial reduction in shots for Bordeaux. As for the biggest negative adjustment in corner kicks, it occurred for Paris Saint-Germain (PSG) in their 1-2 loss to Stade Rennais. PSG trailed for most of the game and failed to capitalize on their opponent's two red cards in the early stages of the second half. Sixteen of their corner kicks came while leading by at least one man but trailing by at least one goal, including 11 corners when up by at least two men.

A notable feature for the negative shifts is that all ten teams (five for shot attempts and five for corner kicks) spent at least half of the game with a manpower advantage, and six of these teams even had a two-man advantage at certain points. This led to our adjustment imposing an even heavier penalty on offensive outputs accumulated during such favorable conditions. In addition to the previously mentioned cases of Bordeaux and PSG attempting most of their shots when up by two men, we also have: Blackpool with 25 of their 26 shots when up by two men, Bournemouth with 7 out of 9 corner kicks, and Getafe with 9 out of 22 shots and 7 out of 10 corner kicks in their home game against Deportivo during the 2009/10 season.

Please note that, despite the multiplicative nature of the projection mechanism, the strength of the adjustment in Tables \ref{tab:StatAdjustShots} and \ref{tab:StatAdjustCorners} was measured by the absolute difference between actual and adjusted values, rather than by percentage change. As a result, the tables primarily highlight teams that both faced some extenuating circumstances \textit{and} had high actual outputs (e.g., 40+ shots for Real Madrid, Inter Milan, and AS Roma, or 15+ corner kicks for AS Roma and PSG). There may have been other teams that experienced higher percentage adjustments but smaller absolute shifts due to lower actual outputs.

Finally, although the home-field factor had a smaller impact on offensive production compared to score and red card differentials, it still influenced the adjustments by awarding extra credit to away teams and applying downward adjustments to home teams. This is partially reflected in the fact that all ten of the strongest negative adjustments occurred for teams playing at home, and seven out of ten strongest positive adjustments happened for teams playing away (for three other teams, the home-field effect was likely overshadowed by the more dominant influences of score and red card differential).

\subsubsection{Season-Long Adjustment}

Beyond the most drastic adjustments observed within individual games, in Tables \ref{tab: Top3_Seasons_Shots} and \ref{tab: Top3_Seasons_Corners} we present some of the strongest positive and negative shifts in season-long average offensive production—measured by shot attempts and corner kicks, respectively—across the 2008–2023 period for each league under consideration.

\begin{table}[]
\centering
\resizebox{\textwidth}{!}{
\begin{tabular}{|l|l|l|l|l|l|l|l|l|}
League & Team & Season & \multicolumn{2}{c}{Shots Per Game (Rank)} & \multicolumn{4}{c}{Actual Shots Per Game (\& Rank) When} \\
 &  &  &   &  & Up & Down & Up & Down \\
 &  &   & Actual & Adjusted, [95\% CI] & $1+$ goal & $1+$ goal & $1+$ man & $1+$ man \\
 
\midrule

\multicolumn{3}{c}{\textbf{Strongest Positive Shifts in Each League:}} & & & & & & \\
ENG & Man City & 2022/23 & 15.7 (\#3) & 17.9, [17.6, 18.1] (\#1) & 7.1 (\#1) & 1.5 (\#20) & 0.3 (\#8) & 0.3 (\#3) \\  % => won the league

GER & Dortmund & 2012/13 & 15.9 (\#3) & 18.0, [17.7, 18.3] (\#2) & 7.7 (\#2) & 2.8 (\#13) & 0.4 (\#10) & 0.9 (\#1) \\ %  => FINISHED #2

FRA & Paris St-Germ & 2017/18 & 16.1 (\#2) & 19.3, [18.9, 19.6] (\#1) & 8.2 (\#1) & 1.7 (\#20) & 0.5 (\#10) & 0.8 (\#1) \\ %  => won the league
 %   => won the league

ITA & Inter Milan & 2008/09 & 14.7 (\#7) & 16.7, [16.5, 16.9] (\#3) & 6.1 (\#1) & 0.6 (\#20) & 0.5 (\#13) & 0.5 (\#8) \\ %  => won the league

SPA & Barcelona & 2009/10 & 15.9 (\#3) & 19.5, [19.1, 19.8] (\#2) & 8.6 (\#2) & 0.7 (\#20) & 0.4 (\#17) & 0.6 (\#3) \\

%... & ... &  ... & ... & ... & ... & ... & ... & ... \\

\multicolumn{3}{c}{\textbf{Strongest Negative Shifts in Each League:}} & & & & & & \\
ENG & Bournemouth & 2018/19 & 11.7 (\#12) & 11.3, [11.2, 11.5] (\#16) & 2.3 (\#13) & 4.5 (\#5) & 1.2 (\#2) & 0.0 (\#18) \\ % => 14th/20

GER & Hannover 96 & 2015/16 & 11.2 (\#13) & 10.9, [10.8, 11.1] (\#16) & 0.9 (\#18) & 6.0 (\#1) & 0.4 (\#6) & 0.0 (\#15)  \\ %=> dead-last (18/18)

FRA & Lorient & 2021/22 & 11.7 (\#10) & 11.1, [10.9, 11.2] (\#15) & 1.1 (\#20) & 4.3 (\#5) & 1.6 (\#1) & 0.2 (\#16)  \\ %=> 16/20

ITA & US Pescara & 2016/17 & 11.5 (\#12) & 10.3, [10.2, 10.5] (\#17) & 0.9 (\#20) & 6.4 (\#1) & 1.2 (\#1) & 0.2 (\#10) \\ % => dead-last (20/20)

SPA & Las Palmas & 2017/18 & 10.7 (\#15) & 10.4, [10.3, 10.5] (\#20) & 1.1 (\#17) & 4.6 (\#3) & 0.7 (\#6) & 0.1 (\#17)\\ % => next-to-last (19/20)

\end{tabular}
}

\caption{Individual seasons across 2008-2023 time span for each of the five European soccer leagues under consideration where a team experienced one of the strongest positive (first five rows) and strongest negative (last five rows) absolute shifts in their shot production per game due to our adjustment. Statistics and league rankings in several important contextual scenarios (leading/trailing, up/down in men) are also provided.
}
\label{tab: Top3_Seasons_Shots}
\end{table}

\begin{table}[]
\centering
\resizebox{\textwidth}{!}{
\begin{tabular}{|l|l|l|l|l|l|l|l|l|}
League & Team & Season & \multicolumn{2}{c}{Corners Per Game (Rank)} & \multicolumn{4}{c}{Actual Corners Per Game (\& Rank) When} \\
 &  &  &   &  & Up & Down & Up & Down \\
 &  &   & Actual & Adjusted, [95\% CI] & $1+$ goal & $1+$ goal & $1+$ man & $1+$ man \\
 
\midrule
\multicolumn{3}{c}{\textbf{Strongest Positive Shifts in Each League:}} & & & & & & \\
ENG & Man City & 2020/21 & 6.4 (\#2) & 7.6, [7.4, 7.9] (\#1) & 2.9 (\#1) & 0.8 (\#20) & 0.2 (\#10) & 0.1 (\#6) \\ % => 1st/20

GER & Dortmund & 2012/13 & 5.5 (\#4) & 6.3, [6.1, 6.5] (\#3) & 2.4 (\#2) & 1.0 (\#17) & 0.1 (\#8) & 0.3 (\#1) \\ %  => 2nd/18

FRA & Paris St-Germ & 2014/15 & 4.5 (\#12) & 5.6, [5.4, 5.7] (\#4) & 2.1 (\#1) & 0.4 (\#20) & 0.0 (\#19) & 0.0 (\#19) \\ % => 1st/20

ITA & Juventus & 2013/14 & 5.7 (\#4) & 7.1, [6.9, 7.3] (\#1) & 2.5 (\#1) & 0.4 (\#19) & 0.2 (\#12) & 0.1 (\#10) \\ %  => 1st/20

SPA & Barcelona & 2012/13 & 5.9 (\#7) & 7.8, [7.5, 8.1] (\#3) & 3.3 (\#1) & 0.5 (\#19) & 0.3 (\#9) & 0.0 (\#18) \\ % => 1st/20

%... & ... &  ... & ... & ... & ... & ... & ... & ... \\

\multicolumn{3}{c}{\textbf{Strongest Negative Shifts in Each League:}} & & & & & & \\
ENG & Aston Villa & 2015/16 & 4.4 (\#15) & 4.2, [4.1, 4.3] (\#19) & 0.3 (\#20) & 2.3 (\#2) & 0 (\#16) & 0.1 (\#12) \\ % => dead-last
GER & FC Cologne & 2017/18 & 5.1 (\#6) & 4.8, [4.7, 5.0] (\#11) & 0.5 (\#18) & 2.7 (\#1) & 0.0 (\#15) & 0.2 (\#8) \\ %=> dead-last/18
FRA & Valenciennes & 2013/14 & 6.2 (\#1) & 6.0, [5.9, 6.2] (\#5) & 0.9 (\#9) & 2.9 (\#1) & 0.4 (\#3) & 0.0 (\#11) \\ % => 19th/20
ITA & Siena & 2009/10 & 5.7 (\#6) & 5.3, [5.2, 5.4] (\#14) & 0.3 (\#20) & 2.8 (\#1) & 0.4 (\#2) & 0.1 (\#14) \\ % => 19th/20
SPA & Recreativo H & 2008/09 & 5.3 (\#9) & 5.1, [5.0, 5.2] (\#16) & 0.4 (\#20) & 2.3 (\#1) & 0.6 (\#3) & 0.0 (\#19) \\ %  => dead-last/20

\end{tabular}
}

\caption{Individual seasons across 2008-2023 time span for each of the five European soccer leagues under consideration where a team experienced one of the strongest positive (first five rows) and strongest negative (last five rows) absolute shifts in their corner kicks per game due to our adjustment. Statistics and league rankings in several important contextual scenarios (leading/trailing, up/down in men) are also provided.
}
\label{tab: Top3_Seasons_Corners}
\end{table}

One may notice that the strongest positive adjustments generally apply to the top-performing teams. Besides passing a domain knowledge sanity check— seeing world-renowned dominant franchises such as Manchester City, Borussia Dortmund, Barcelona, Paris Saint-Germain, Juventus, and Inter Milan—several other key observations reinforce this pattern. First, these teams consistently generate high offensive output while in the lead, i.e., up by at least one goal. Tables \ref{tab: Top3_Seasons_Shots} and \ref{tab: Top3_Seasons_Corners} show all of them ranking \#1 or \#2 in that metric within their respective leagues for the given season. This reflects both the significant time they spend in the lead and their ability to maintain high offensive production while ahead—common traits of dominant teams. Second, these teams typically rank near the bottom in terms of shots taken while trailing (often \#18–\#20 out of 20 teams in most leagues, or \#17–\#18 out of 18 teams in the German Bundesliga), indicating how infrequently they fall behind—another hallmark of strong performance. Although not shown in the tables, we verified that teams with the largest positive adjustments also spent the least amount of time per game trailing (or, at worst, the third lowest), and the most time in the lead (or, at worst, the third highest). Conversely, the largest negative adjustments correspond to teams that consistently rank low in output generated while in the lead, instead producing the majority of their offensive output while trailing—when the game context is most favorable for such production.

On the other hand, identifying consistent season-long patterns related to playing with a numerical advantage or disadvantage is more challenging. One could argue that, for shot attempts in particular (Table \ref{tab: Top3_Seasons_Shots}), there is a tendency for teams with the strongest positive adjustments to rank highly in shots generated while at a man disadvantage (most appear in the top three). Conversely, teams with negative adjustments tend to rank low in that same metric (typically \#15 or lower out of 20 teams), while consistently ranking high in shots produced when up at least one man. That said, these patterns are less pronounced than in the case of score differentials, and there is no clear trend among positively adjusted teams when it comes to generating shots while holding a 1-man advantage. As for corner kicks (Table \ref{tab: Top3_Seasons_Corners}), there are no discernible patterns in teams’ season-long tendencies related to playing with a numerical advantage or disadvantage.

This contrast between the impact of score differential and red card differential (i.e., playing with a 1-man advantage or disadvantage) is not surprising for one main reason: red card situations are generally much rarer than leading or trailing, as goals occur more frequently than player dismissals. Consequently, unlike scoring contexts, it is harder for red card situations to have a sustained impact over the course of an entire season. This contrasts with individual games where, as seen in Tables \ref{tab:StatAdjustShots} and \ref{tab:StatAdjustCorners}, red cards were present in nearly all matches most affected by our contextual adjustment.

It is also noteworthy to examine changes in league rankings for the offensive output under consideration (shots in Table \ref{tab: Top3_Seasons_Shots}, corners in Table \ref{tab: Top3_Seasons_Corners}) before and after the adjustment. Note that, for these season-long tables, we specifically selected the strongest positive and negative adjustments that resulted in at least some movement up or down the rankings within that season (there were teams experiencing larger absolute adjustments, but with their ranking being unaffected). % for teams that remained ranked \#1 in both unadjusted and adjusted rankings).
In eight out of ten positive season-long adjustments across both tables, the team went on to win the league that season, with Dortmund's 2012/13 campaign being the only second-place finish, albeit reflected twice (for both shots and corner kicks). These league-winners include teams like PSG, who originally ranked only \#12 in corner kicks per game and rose to \#4 post-adjustment, and Inter Milan, who moved from \#7 to \#3 in shot attempts per game. Conversely, eight out of ten negative season-long adjustments involved teams that finished in the bottom two of their respective league, and therefore got relegated to a lower-level league. For example, Serie A’s US Pescara initially ranked \#12 (out of 20) with 11.5 shot attempts per game, but their adjusted rate dropped to 10.3 (95\% CI: [10.2, 10.5]), lowering them to \#17—more accurately reflecting their frequent trailing game states and aligning better with their final standing. Even more notably, teams like Valenciennes, FC Cologne, and Siena, despite finishing in the bottom two of their league, initially ranked very highly in corner kicks per game; our adjustments substantially lowered their respective rankings in that statistic.

Tables \ref{tab:CorrAnalysisShots} and \ref{tab:CorrAnalysisCorners} provide a more formal summary of the season-long patterns discussed above. For both shots and corner kicks, one can observe a much stronger positive correlation between the adjusted statistics and the points earned, indicating better alignment with the final standings. Moreover, consistent with the intuition behind our adjustment mechanism, there is a stronger positive relationship between total shots/corners per game and shots/corners taken while leading in score (a setting more conducive to offensive production), and a stronger negative association with shots/corners taken while trailing (a less conducive setting, respectively). In contrast, the correlation patterns between offensive outputs and situations involving a manpower advantage or disadvantage are less clear. First, these correlations are much weaker than those related to scoring context—typically ranging from 0.05 to 0.20 in magnitude, compared to 0.25 to 0.90 for when a team is up or down by one or more goals. Second, the higher standard errors associated with these manpower-related correlations indicate a lack of consistency, resulting in inconclusive comparisons between actual and adjusted statistics. These findings support our earlier takeaways regarding season-long adjustment patterns: scoring context drives most of the observed shifts, as it reflects a more sustainable dynamic over the course of a season (i.e., stronger teams tend to lead more often, while weaker teams tend to trail). In contrast, manpower advantages or disadvantages play a less prominent role, due to the irregular and less predictable nature of red card occurrences.

\begin{table}[h!]
    \centering
    \begin{minipage}{0.45\textwidth}
        \centering
  \begin{tabular}{lcc}
    \hline
     & \multicolumn{2}{|c|} {Correlation (SE) With}    \\
    Statistic (Per Game) & Actual Shots &  Adjusted Shots \\
    \midrule
    Points Earned & 0.718 (0.01) & 0.787 (0.01)  \\
    Shots Up 1+ Goal & 0.829 (0.01) & 0.894 (0.01) \\
    Shots Down 1+ Goal & -0.254 (0.02) \ & -0.370 (0.02) \  \\
    Shots Up 1+ Men  & 0.190 (0.03) & 0.129 (0.03)  \\
    Shots Down 1+ Men & 0.051 (0.02) & 0.089 (0.02)  \\
    \hline
  \end{tabular}
        \caption{Correlation of several statistical categories with the actual and adjusted shots. Averaged across five leagues and 15 seasons under consideration, with the standard error provided in parentheses.}
        \label{tab:CorrAnalysisShots}
    \end{minipage}%
    \hfill % this adds space between the tables
    \begin{minipage}{0.45\textwidth}
  \begin{tabular}{lcc}
    \hline
     & \multicolumn{2}{|c|} {Correlation (SE) With}    \\
    Statistic (Per Game) & Actual Corners &  Adjusted Corners \\
    \midrule
    Points Earned &  0.603 (0.02) & 0.737 (0.01) \\
    Corners Up 1+ Goal &  0.714 (0.02) & 0.849 (0.01) \\
    Corners Down 1+ Goal &  -0.057 (0.03) \ & -0.252 (0.03) \ \\
    Corners Up 1+ Men   & 0.151 (0.03) & 0.093 (0.03) \\
    Corners Down 1+ Men  & 0.059 (0.03) & 0.090 (0.03) \\
    \hline
  \end{tabular}
        \caption{Correlation of several statistical categories with the actual and adjusted corners. Averaged across five leagues and 15 seasons under consideration, with the standard error provided in parentheses.}
        \label{tab:CorrAnalysisCorners}
    \end{minipage}
   % \caption{Correlation of several statistical categories with the actual and adjusted shots (5a) and corners (5b), respectively. Averaged across five leagues and 15 seasons under consideration, with the standard error provided in parentheses.}
\end{table}

All in all, we observe a pattern where the adjusted numbers align more closely with the actual final standings. This results from accounting for the game context in which offensive outputs are accumulated, thereby reducing the skew in statistics that may arise purely from tactical choices rather than from genuinely outperforming the opponent. As illustrated, certain teams may at times display misleadingly high or low values in metrics such as shot attempts and corner kicks, potentially resulting in misinterpretation.

\subsubsection{Forecasting performance comparison}

\begin{figure}[]
   \centering
    \includegraphics[scale=0.65]{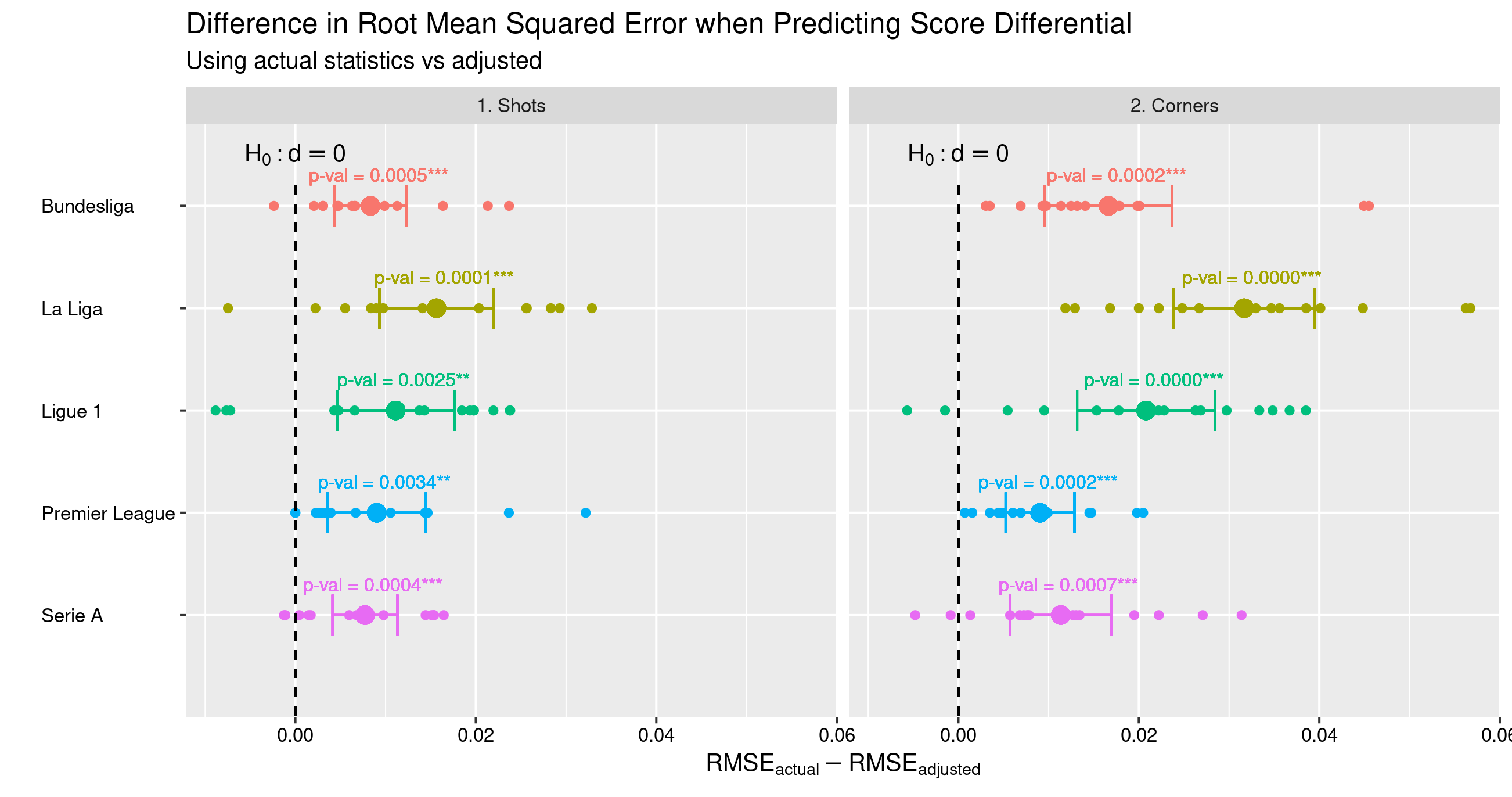}
   
    \caption{%Comparing the forecasting performance via linear regression that uses home and away team's actual and adjusted statistics, respectively. 
    Paired $t$-test $p$-values and 95\% confidence intervals for the difference in root mean squared error (RMSE) when predicting final score differentials of future games within each season (individual points) and each league (y-axis) under consideration. Forecasts were generated using linear regression with two explanatory variables: the home and away teams' statistics (shots or corners) up to the midpoint of the season, either left unchanged ("actual") or adjusted using our proposed mechanism ("adjusted").
    % (shots actual and adjusted statistics, respectively.}
    }
    
    \label{fig:ForecastPerformance}
\end{figure}

Figure \ref{fig:ForecastPerformance} presents the results of the paired $t$-test comparing forecasting performance when using adjusted statistics versus actual values in predicting score differentials.

One may notice that forecasts based on actual statistic values consistently result in higher RMSE compared to those based on adjusted values, with all Holm-adjusted $p$-values significant at the 0.01 level—and some even at the 0.001 level. That said, the confidence intervals for the true RMSE difference range from as low as 0.004 goals per game (English Premier League, shots) to only as high as 0.04 goals per game (Spanish La Liga, corners). When projected over a 380-game schedule—which corresponds to a single season in a 20-team league—this translates to a difference of just 1.52 to 15.2 total goals over a season, which may not be substantial from a practical standpoint. Furthermore, the absolute RMSE values for each season typically ranged between 1.3 and 1.9 goals per game—unsurprisingly high, given the simplicity of the linear regression model used for forecasting—suggesting that the observed per-game RMSE reductions of 0.004–0.04 are of marginal practical magnitude. Nonetheless, when considered alongside the statistically significant results discussed above, these findings still provide evidence of a stable discernible improvement that our adjustment provides.

\section{Conclusion \& Future Work}
    \label{sec:ConclusionFutureWork}

We modeled teams' offensive production—such as shot attempts and corner kicks—as a function of various factors that influence the likelihood of employing certain tactics. The model that achieved the best balance between goodness-of-fit and out-of-sample predictive performance was Negative Binomial Generalized Additive Model (GAM) with smoothing splines, in which we binned extreme values for score differential (treating all differentials beyond $+2$ the same, and beyond $-2$, respectively), red card differential (beyond $+1$ and $-1$, respectively), and game minute (beyond 45, i.e., extra time). We also confirmed via hypothesis testing procedures that each variable in the baseline set of five factors—score differential, red card differential, home indicator, win probability differential, and game minute (including a half indicator)—was statistically significant and should be retained.

The nature of the effects observed in our analysis was largely intuitive. Notably, the impact of score differential on offensive output exhibited a predictable decreasing pattern as the differential gradually shifted from $-3$ to $3$. The influence of red card differential was even more pronounced in terms of practical significance, revealing a strong negative trend as the differential increased. These effects align with the understanding that teams in the lead tend to adopt a more conservative approach to protect their advantage and secure the valuable three points for a win, whereas teams down by at least one player are often forced to play defensively due to their numerical disadvantage. The pre-match win probability differential showed a strong positive association with offensive output, reinforcing the notion that stronger teams tend to outperform their opponents. Additionally, the home factor consistently had a positive, albeit relatively modest, effect on a team’s offensive production. All of these findings are consistent with existing literature, including [X, \citep{lopez2018often, harville1977use}].

Incorporating the game minute variable directly into the model allowed us to explore several research questions that would not have been possible with aggregate data alone. First, it helped illustrate trends in offensive output as the half progresses. Specifically, it showed a steady increase in offensive activity during the first 5 minutes of a half, with the starting level of activity being higher in the second half. This could be attributed to both teams calibrating their respective offenses for the opponent early in the first half, and getting back into the flow of the game early in the second half, albeit already having calibrated their offense (hence the higher baseline activity level). Afterward, the offensive activity level remained relatively stable. Secondly, we used the game minute variable to test several hypotheses driven by domain knowledge and intuition. For example, one might expect that as time expires, the trailing team becomes more desperate to try and score, while the leading team becomes more conservative in order to hold on to the lead. This would be captured by an interaction between score differential and game minute. When testing the significance of this interaction, we found that there is very limited evidence to support this hypothesis. In the few cases where the interaction was found statistically significant, the results did align with the intuition, but such cases were rare (out of 75 tests). We also tested interactions between other key factor variables, as several plausible hypotheses could be explored that way. However, we found even less evidence for these effects than for the interaction between game minute and score differential.

In our statistical adjustment process, we projected each team's offensive performance in a selected statistical category onto a baseline scenario of a tied home game at even strength, aiming to provide a fairer representation of offensive production given the game context. This adjustment involved giving greater weight to statistics accumulated during periods when a team was leading and/or playing with fewer men, as these situations are generally less favorable for offensive output. Conversely, statistics were adjusted downward for periods when a team was trailing or had a numerical advantage. When examining the largest adjustments to team performances—both in individual games and season-long per-game averages—we observed a clear tendency for teams to be rewarded for generating significant offensive output while leading by at least one goal. In contrast, uneven player distributions (e.g., 9 or 10 players vs. 11) played a more prominent role at the single-game level, as it is uncommon for teams to experience such scenarios consistently over an entire season. Nonetheless, some season-long patterns of offensive outputs generated while up/down in men were still identifiable, just not as consistently as when it came to the score differential scenarios.

Note that our statistical adjustment aimed to make team performance comparisons in categories like shot attempts and corner kicks more reflective of relative levels of play. For example, in season-long adjustments, it helped align team rankings in shot attempts and corner kicks more closely with their final standings by points—benefiting stronger teams and downgrading weaker ones. This adjustment accounted for game context, where top teams tend to lead and bottom teams tend to trail for most of the time, thus correcting for potential skew in raw game totals. Notably, it addressed cases where bottom-two teams by points ranked near the top of the league in corner kicks per game, which could otherwise be misleading. Moreover, as a more objective way to quantify the improvement gained from our adjustment, we demonstrated that team per-game averages based on adjusted statistics outperform those based on actual statistics when used to forecast final score differentials of future games. Although not overly large from a practical standpoint, the improvement was consistent enough to be deemed statistically significant, with the findings replicated across all five leagues under consideration.

As for the single-game adjustments, they also tended to align well with the final game outcome—teams with positive adjustments generally won, while those with negative adjustments tended to lose. Beyond merely reflecting the tendency to play with a lead or deficit, these cases often involved more extreme circumstances, such as teams gaining a 1- or 2-man advantage (or suffering a corresponding disadvantage). For example, Getafe lost 0–2 at home to Deportivo in their 2009/10 match despite outshooting them 22–8 and earning 14 corner kicks to the opponent's 3. However, Getafe played with at least a 1-man advantage for most of the game, and a 2-man advantage for the final 30+ minutes, skewing the raw numbers. Our adjustment reduced their shot count to 12.8 and corners to 8.7, while Deportivo’s adjusted figures rose to 14.6 and 7.8, respectively—offering a more context-aware comparison that accounts for Deportivo’s shorthanded status while also holding a lead for most of the game. Similar adjustments occurred in other matches:
\begin{itemize}
    \item Stade Rennais' 2–1 win over Paris Saint-Germain in 2012/13, where PSG led 23–8 in shots and 18–3 in corners, but the adjustment reversed shots to 12.3–16.2, while shrinking the corner difference (9.5–5.9).
    \item Montpellier's 2–0 win over Bordeaux in 2021/22, despite trailing 32–7 in shots and 11–4 in corners; adjustments brought the numbers to 13.8–11.5 and 5.0–6.1, respectively.
    \item Bayern Munich’s 4–0 win against Stuttgart in 2020/21, which initially showed just a slight 15–12 edge in shots and a 1–3 deficit in corners. After adjustment, shots became 31.8–9.6 and corners — 1.7–2.5.
\end{itemize}
These examples highlight the main goal of our adjustment, which is to make raw match statistics like shot attempts and corner kicks less misleading and more reflective of the relative quality of play. Without such context, one might incorrectly attribute a match outcome solely to luck, ignoring the fact that such statistical imbalance is often due to tactical dynamics driven by game state.

% https://www.espn.com/soccer/matchstats/_/gameId/576779
% https://www.espn.com/soccer/matchstats/_/gameId/342835
% https://www.espn.com/soccer/matchstats/_/gameId/275872
% https://www.espn.com/soccer/matchstats/_/gameId/609246

% 15-12, 1-3, to 31.8 vs 9.6, 1.71 vs 2.48

% (For PSG vs Stade: going from 23-8, 18-3 to 12.3-16.2, 9.5 vs 5.9; for Bayern vs Stuttgart: going from 15-12, 1-3, to 31.8 vs 9.6, 1.71 vs 2.48; for Getafe-Deportivo: 12.8 vs 14.6, 8.7 vs 7.8) 
% 576779
% Shots: 31.8 vs 9.6
% Corners: 1.71 vs 2.48)

Returning to the main motivating example of the 2022 FIFA World Cup Quarterfinal between France and England: even when accounting for France having spent a significant portion of the game with a 1-goal lead and adopting a more defensive style, the extent to which England numerically outperformed France during that period—11 to 1 in shots, 5 to 0 in corner kicks—cannot be fully explained by contextual factors alone. Applying even the most extreme adjustment for leading or trailing by one goal (see Figure \ref{fig:MultCoef_BINNED}), would project England to have 14.35 shots ($5 + 11\times0.85$) instead of their actual 16, and France to have 8.15 shots ($7 + 1 \times 1.15$) instead of their actual 8. This reduces England’s statistical lead by only about two shots. The fact that England’s lead in shots remains substantial even after adjustment, in our view, actually strengthens the argument that, despite having lost the game, they might have played more dominantly and might have gotten unlucky. This illustrates that even when our adjustment does not drastically alter the relative performance of the teams, it still provides a fairer baseline for comparison—enabling more confident conclusions about which team truly controlled the game.

For future work, a promising direction would be to model the actual quality of shot attempts generated in various game contexts, as not all shots are equally likely to pose a significant threat to the opponent. To this end, we plan to use expected goal values ($xG$) \citep{whitmore_what_xG_2023} as our measure of offensive output. This would allow us to analyze how the quality of scoring opportunities is influenced by game context, with the potential for subsequent statistical adjustments. Whether a similar adjustment approach is appropriate in this setting will depend heavily on the presence of a consistent monotonic relationship between game context and $xG$ output, ensuring that the adjustment is meaningful. For example, projecting performance onto a neutral context (e.g., tied score or even strength) is only valid if shorthanded teams with a lead consistently generate lower $xG$ values, and vice versa for teams that trail and/or have manpower advantage.

Lastly, it is important to clarify that, although we referenced a World Cup elimination game as an example, this study primarily focuses on national club leagues, where elimination games—at least in the same sense as the World Cup's knockout stages—do not exist. In these leagues, points are awarded based on match outcomes: 3 points for a win, 1 point for a draw, and 0 points for a loss. The league winner is determined by the total number of points accumulated over the course of the season. With this context in mind, examining competitions that rely on binary outcomes (win/loss) or score differentials to determine advancement would be a logical extension of this research. Such competitions include not only the World Cup knockout rounds, but also the elimination stages of the Champions League, Europa League, and tournaments like the Football Association Challenge Cup (FA Cup).

\section{Acknowledgments}

\if0\blind
{
The authors are grateful to New College of Florida for providing summer research funding. The authors used ChatGPT to edit the text for clarity, grammar and syntax, but made sure to subsequently review the text themselves and confirm the actual meaning was preserved.
} \fi

\if1\blind
{
The authors are grateful to the host institution for providing summer research funding. The authors used ChatGPT to edit the text for clarity, grammar and syntax, but made sure to subsequently review the text themselves and confirm the actual meaning was preserved.
} \fi

\section{Declaration of interest statement}

Authors of this work confirm that there are no known conflicts of interest to disclose.

\if0\blind
{
\subsection*{Data and code availability}
Data and source code for this work have been made publicly available on Github via this link: \url{https://github.com/UsDAnDreS/JQAS_LeveragingMinuteByMinute_Soccer_Event_Data_Paper}.
} \fi

%\printbibliography
\bibliographystyle{plainnat}  % Choose the bibliography style (plainnat, unsrtnat, etc.)
\bibliography{bibliography}  % Point to your .bib file here

\begin{thebibliography}{27}
\providecommand{\natexlab}[1]{#1}
\providecommand{\url}[1]{\texttt{#1}}
\expandafter\ifx\csname urlstyle\endcsname\relax
  \providecommand{\doi}[1]{doi: #1}\else
  \providecommand{\doi}{doi: \begingroup \urlstyle{rm}\Url}\fi

\bibitem[esp()]{espnFranceEngland}
{ESPN: France 2-1 England Commentary}.
\newblock \url{https://www.espn.com/soccer/commentary/_/gameId/633846}.

\bibitem[noa()]{noauthor_selenium_nodate}
{The Selenium Browser Automation Documentation}.
\newblock \url{https://www.selenium.dev/documentation/}.

\bibitem[Akaike(1998)]{akaike1998information}
Hirotogu Akaike.
\newblock Information theory and an extension of the maximum likelihood principle.
\newblock In \emph{Selected papers of hirotugu akaike}, pages 199--213. Springer, 1998.

\bibitem[Boshnakov et~al.(2017)Boshnakov, Kharrat, and McHale]{boshnakov2017bivariate}
Georgi Boshnakov, Tarak Kharrat, and Ian~G McHale.
\newblock A bivariate weibull count model for forecasting association football scores.
\newblock \emph{International Journal of Forecasting}, 33\penalty0 (2):\penalty0 458--466, 2017.

\bibitem[Cefis and Carpita(2025)]{cefis2025new}
Mattia Cefis and Maurizio Carpita.
\newblock A new xg model for football analytics.
\newblock \emph{Journal of the Operational Research Society}, 76\penalty0 (1):\penalty0 1--13, 2025.

\bibitem[Cefis et~al.(2022)]{cefis2022football}
Mattia Cefis et~al.
\newblock Football analytics: a bibliometric study about the last decade contributions.
\newblock \emph{Electronic Journal of Applied Statistical Analysis}, 15\penalty0 (1):\penalty0 232--248, 2022.

\bibitem[Cemek et~al.(2025)Cemek, Skripnikov, and Gillman]{cemek2025statistical}
Ahmet Cemek, Andrey Skripnikov, and David Gillman.
\newblock Statistical adjustment for tactical choices when evaluating team’s offensive output across five major european club soccer leagues.
\newblock 2025.

\bibitem[Edwards and Archambault(1979)]{edwards1979home}
John Edwards and Denise Archambault.
\newblock The home field advantage.
\newblock \emph{Sports, games, and play: Social and psychological viewpoints}, pages 409--438, 1979.

\bibitem[FBref()]{fbref}
FBref.
\newblock {Expected Goals (xG) Model Explained}.
\newblock \url{https://fbref.com/en/expected-goals-model-explained/}.

\bibitem[Hartig and Hartig(2017)]{hartig2017package}
Florian Hartig and Maintainer~Florian Hartig.
\newblock Package ‘dharma’.
\newblock \emph{R package}, 531:\penalty0 532, 2017.

\bibitem[Harville(1977)]{harville1977use}
David Harville.
\newblock The use of linear-model methodology to rate high school or college football teams.
\newblock \emph{Journal of the American Statistical Association}, 72\penalty0 (358):\penalty0 278--289, 1977.

\bibitem[Holm(1979)]{holm1979simple}
Sture Holm.
\newblock A simple sequentially rejective multiple test procedure.
\newblock \emph{Scandinavian journal of statistics}, pages 65--70, 1979.

\bibitem[Kempton et~al.(2016)Kempton, Kennedy, and Coutts]{kempton2016expected}
Thomas Kempton, Nicholas Kennedy, and Aaron~J Coutts.
\newblock The expected value of possession in professional rugby league match-play.
\newblock \emph{Journal of sports sciences}, 34\penalty0 (7):\penalty0 645--650, 2016.

\bibitem[Lopez et~al.(2018)Lopez, Matthews, and Baumer]{lopez2018often}
Michael~J Lopez, Gregory~J Matthews, and Benjamin~S Baumer.
\newblock How often does the best team win? a unified approach to understanding randomness in north american sport.
\newblock \emph{The Annals of Applied Statistics}, 12\penalty0 (4):\penalty0 2483--2516, 2018.

\bibitem[Macdonald(2012)]{macdonald2012expected}
Brian Macdonald.
\newblock An expected goals model for evaluating nhl teams and players.
\newblock In \emph{Proceedings of the 2012 mit sloan sports analytics conference}, 2012.

\bibitem[McCullagh(2019)]{mccullagh2019generalized}
Peter McCullagh.
\newblock \emph{Generalized linear models}.
\newblock Routledge, 2019.

\bibitem[Mead et~al.()Mead, O’Hare, and {McMenemy}]{expectedGoal}
James Mead, Anthony O’Hare, and Paul {McMenemy}.
\newblock Expected goals in football: Improving model performance and demonstrating value.
\newblock 18\penalty0 (4):\penalty0 e0282295.
\newblock ISSN 1932-6203.
\newblock \doi{10.1371/journal.pone.0282295}.
\newblock URL \url{https://journals.plos.org/plosone/article?id=10.1371/journal.pone.0282295}.
\newblock Publisher: Public Library of Science.

\bibitem[Schwarz(1978)]{schwarz1978estimating}
Gideon Schwarz.
\newblock Estimating the dimension of a model.
\newblock \emph{The annals of statistics}, pages 461--464, 1978.

\bibitem[Shapiro and Wilk(1965)]{shapiro1965analysis}
Samuel~Sanford Shapiro and Martin~B Wilk.
\newblock An analysis of variance test for normality (complete samples).
\newblock \emph{Biometrika}, 52\penalty0 (3-4):\penalty0 591--611, 1965.

\bibitem[Sosa(2015)]{sosa2015identidad}
Francisco Gabriel~Ruiz Sosa.
\newblock La identidad del italiano en la evoluci{\'o}n del catenaccio.
\newblock \emph{Impetus}, 9\penalty0 (2):\penalty0 135--142, 2015.

\bibitem[StatsBomb({\natexlab{a}})]{StatsBomb_ExpectedPass}
StatsBomb.
\newblock xpass 360: Upgrading expected pass models.
\newblock \url{https://statsbomb.com/articles/soccer/xpass-360-upgrading-expected-pass-xpass-models/}, {\natexlab{a}}.

\bibitem[StatsBomb({\natexlab{b}})]{StatsBomb_PossessionValue}
StatsBomb.
\newblock Examples of possession value models.
\newblock \url{https://statsbomb.com/soccer-metrics/possession-value-models-explained/}, {\natexlab{b}}.

\bibitem[Trequattrini et~al.(2016)Trequattrini, Del~Giudice, Cuozzo, and Palmaccio]{trequattrini2016does}
Raffaele Trequattrini, Manlio Del~Giudice, Benedetta Cuozzo, and Matteo Palmaccio.
\newblock Does sport innovation create value? the case of professional football clubs.
\newblock \emph{Technology, Innovation and Education}, 2:\penalty0 1--15, 2016.

\bibitem[Van~Roy et~al.(2020)Van~Roy, Robberechts, Decroos, and Davis]{van2020valuing}
Maaike Van~Roy, Pieter Robberechts, Tom Decroos, and Jesse Davis.
\newblock Valuing on-the-ball actions in soccer: a critical comparison of xt and vaep.
\newblock In \emph{Proceedings of the 2020 AAAI Workshop on AI in Team Sports}, pages 1--8, 2020.

\bibitem[Whitmore()]{whitmore_what_xG_2023}
Jonny Whitmore.
\newblock {What is Expected Goals?}
\newblock \url{https://theanalyst.com/2023/08/what-is-expected-goals-xg/}.

\bibitem[Wood(2017)]{wood2017generalized}
Simon~N Wood.
\newblock \emph{Generalized additive models: an introduction with R}.
\newblock chapman and hall/CRC, 2017.

\bibitem[Yurko et~al.(2019)Yurko, Ventura, and Horowitz]{yurko2019nflwar}
Ronald Yurko, Samuel Ventura, and Maksim Horowitz.
\newblock nflwar: a reproducible method for offensive player evaluation in football.
\newblock \emph{Journal of Quantitative Analysis in Sports}, 15\penalty0 (3):\penalty0 163--183, 2019.

\end{thebibliography}

\section{Appendix}

\subsection{Model comparison: Poisson, Negative Binomial, Zero-Inflated Poisson}

\label{sec:Appendix_Pois_NB_ZIP_AIC_BIC}

Besides the DHARMa diagnostics and leave-one-season-out calibration plots, we also carried out an AIC/BIC comparison of count-response models under consideration.

\begin{table}[h]
\centering
\begin{tabular}{|>{\centering\arraybackslash}m{2cm}|>{\centering\arraybackslash}m{2cm}|>{\centering\arraybackslash}m{2cm}|>{\centering\arraybackslash}m{2cm}|>{\centering\arraybackslash}m{2cm}|>{\centering\arraybackslash}m{2cm}|>{\centering\arraybackslash}m{2cm}|}
% \begin{tabular}{|l|l|l|l|l|l|l|}
\hline
\multicolumn{1}{|c|} {} & \multicolumn{3}{|c|} {Shot Attempts} & \multicolumn{3}{c|}{Corner Kicks} \\
 \hline
% \diagbox{Leagues}{Distributions} & Poisson & Negative Binomial & Zero-Inflated Poisson &  Poisson & Negative Binomial & Zero-Inflated Poisson \\
Leagues & Poisson & Negative Binomial & Zero-Inflated Poisson &  Poisson & Negative Binomial & Zero-Inflated Poisson \\
\midrule
Bundesliga & 695012.5 (74) &  693432.3 (71) & \textbf{693113.7 (106)} &  344286.1 (69) &
343844.3 (68) & \textbf{343792.6 (104)} \\ 

La Liga &  703460.4 (81) &   702608.5 (78) & \textbf{702148.5 (126)} &  373223.2 (71) &  372732.8 (73) &  \textbf{372711.5 (99)} \\

Ligue 1 & 790313.9 (81) & 788846.0 (78) & \textbf{788451.6 (123)} & 406397.3 (66) & 405661.2 (66) & \textbf{405623.6 (94)} \\

Premier League &  722807.5 (87) & 720360.4 (82) & \textbf{719835.4 (124)} & 380695.0 (74) & 380089.1 (75) & \textbf{380022.9 (98)} \\

Serie A & 813704.8 (80) &  812576.1 (76) &  \textbf{811871.9 (118)} & 414359.2 (70) & 413629.8 (68) &  \textbf{413501.0 (99)} \\ 
\hline
\end{tabular}
\caption{AIC values (with model complexity in parentheses) comparing all count-based models under consideration across all five leagues. Best-performing models marked in \textbf{bold}.}
\label{tab:AIC_Pois_NegBin_ZIP}
\end{table}

\begin{table}[ht]
\centering
\begin{tabular}{|>{\centering\arraybackslash}m{2cm}|>{\centering\arraybackslash}m{2cm}|>{\centering\arraybackslash}m{2cm}|>{\centering\arraybackslash}m{2cm}|>{\centering\arraybackslash}m{2cm}|>{\centering\arraybackslash}m{2cm}|>{\centering\arraybackslash}m{2cm}|}
\hline
\multicolumn{1}{|c|} {} & \multicolumn{3}{|c|} {Shot Attempts} & \multicolumn{3}{c|}{Corner Kicks} \\
\hline
% \diagbox{Leagues}{Distributions} & Poisson & Negative Binomial & Zero-Inflated Poisson &  Poisson & Negative Binomial & Zero-Inflated Poisson \\
Leagues & Poisson & Negative Binomial & Zero-Inflated Poisson &  Poisson & Negative Binomial & Zero-Inflated Poisson \\
\midrule
Bundesliga & 695873.0 (79) &  \textbf{694261.8 (71)} &  694352.4 (106) & 345090.0 (69) &  \textbf{344633.0 (68)} & 345006.2 (104) \\

La Liga &   704414.3 (81) &  \textbf{703522.9 (78)} & 703623.9 (126) &  374053.6 (71) &  \textbf{373592.5 (73)} & 373868.1 (99)\\

Ligue 1 &   791269.0 (81) & \textbf{789743.6 (78)} &  789915.3 (124) & 407178.0 (66) &   \textbf{406445.1 (66)} &  406737.4 (94) \\

Premier League &   723826.5 (87) &  721320.0 (82) & \textbf{721287.6 (124)} & 381562.0 (74) & \textbf{380965.2 (75)} & 381258.4 (98) \\

Serie A & 814655.2 (80) &  813475.3 (76) & \textbf{813264.4 (118)} &  415189.4 (70) &  \textbf{414437.7 (68)} & 414665.8 (98) \\
\hline
\end{tabular}
\caption{BIC values (with model complexity in parentheses) comparing all count-based models under consideration across all five leagues. Best-performing models marked in \textbf{bold}.}
\label{tab:BIC_Pois_NegBin_ZIP}
\end{table}

%\newpage

\subsection{Nature of Effects: Initial Smoothing Splines GAM Fit}

\begin{figure}[H]
    \centering
   \includegraphics[scale=0.25]{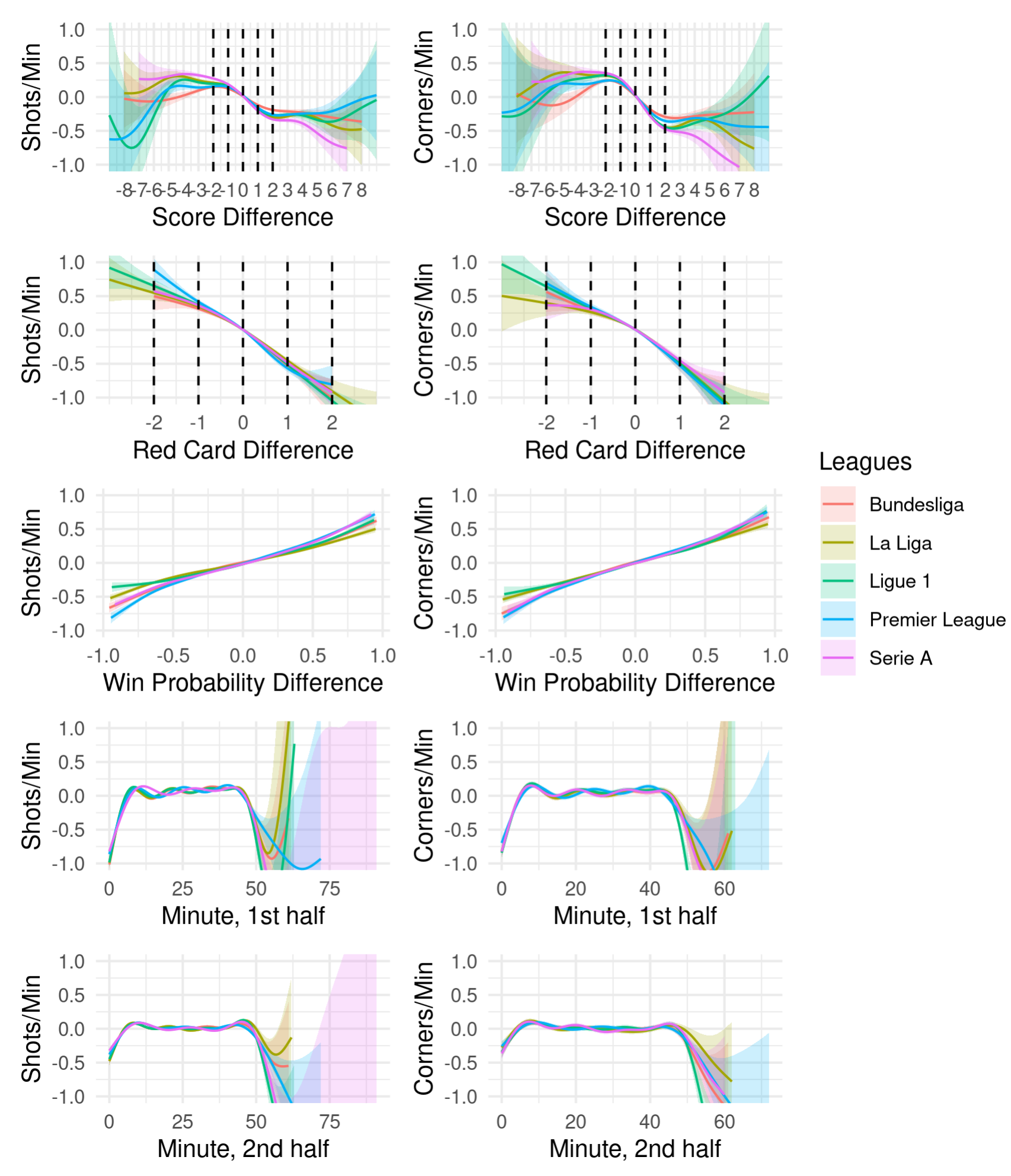}
    \caption{Nature of effects for score differential, red card differential, win probability differential and game minute in 1st/2nd halves on shot attempts (left) and corner kicks (right) in the baseline Negative Binomial Generalized Additive \textbf{smoothing splines} model fitted to 15 seasons (2008-2023) across five major European soccer leagues. The y-axis is on the scale of linear predictor (log-response).}
    \label{fig:EffectDisplays}
\end{figure}

\iffalse

\begin{table}[ht]
  \centering
%\begin{minipage}[t]{0.48\textwidth}

  \centering
  \begin{tabular}{lcccc}
    \hline
     & Correlation (SE) With   & Correlation (SE) With & Correlation (SE) With   & Correlation (SE) With  \\
    Statistic (Per Game) & Actual Shots &  Adjusted Shots & Actual Corners &  Adjusted Corners \\
    \midrule
    Points Earned & 0.718 (0.01) & 0.787 (0.01) & 0.603 (0.02) & 0.737 (0.01) \\
    Shots Up 1+ Goal & 0.829 (0.01) & 0.894 (0.01) & 0.714 (0.02) & 0.849 (0.01) \\
    Shots Down 1+ Goal & -0.254 (0.02) \ & -0.370 (0.02) \ & -0.057 (0.03) \ & -0.252 (0.03) \ \\
    Shots Up 1+ Men  & 0.190 (0.03) & 0.129 (0.03) & 0.151 (0.03) & 0.093 (0.03) \\
    Shots Down 1+ Men & 0.051 (0.02) & 0.089 (0.02) & 0.059 (0.03) & 0.090 (0.03) \\
    \hline
  \end{tabular}
    \caption{Correlation of several statistical categories with the actual and adjusted shots (left) and corners (right), respectively. Averaged across five leagues and 15 seasons under consideration, with the standard error provided in parentheses.}
  \label{tab:CorrAnalysis}
\end{table}

\fi

\end{document}